\documentclass[sigconf,10pt]{acmart}

\usepackage{graphicx} %
\usepackage{xspace}
\usepackage{subcaption}
\usepackage{url}
\usepackage{multirow}
\usepackage{cancel}
\usepackage{hyperref}
\usepackage{algorithmic}
\usepackage{algorithm}

\usepackage[absolute,overlay]{textpos}
\newcommand{\name}{XRLoc\xspace}
\renewcommand{\iota}{\textsl{j}\xspace}

\usepackage{etoolbox}
\apptocmd\normalsize{%
 \abovedisplayskip=2pt plus 2pt minus 2pt
 \abovedisplayshortskip=2pt plus 2pt
 \belowdisplayskip=2pt plus 2pt minus 2pt
 \belowdisplayshortskip=2pt plus 2pt minus 2pt
}{}{}

\usepackage[skip=0pt]{subcaption}
\usepackage[explicit,noindentafter]{titlesec}
\titlespacing\section{3pt}{4pt plus 2pt minus 2pt}{1pt plus 2pt minus 2pt}
\titlespacing\subsection{3pt}{4pt plus 2pt minus 2pt}{1pt plus 2pt minus 2pt}
\titlespacing\subsubsection{3pt}{4pt plus 2pt minus 2pt}{1pt plus 2pt minus 2pt}

\acmDOI{}
\acmISBN{}
\acmConference[]{}
\acmYear{2023}
\copyrightyear{}
\acmPrice{}

\renewcommand\footnotetextcopyrightpermission[1]{} %
\setcopyright{none}

\title{\name: Accurate UWB Localization to \\
Realize XR Deployments}

\author{Aditya Arun}
\authornote{Both authors contributed equally to this research.}
\affiliation{%
  \institution{University of California San Diego}
  \city{California}
  \country{United States}
}
\email{aarun@ucsd.edu}

\author{Shunsuke Saruwatari}
\authornotemark[1]
\affiliation{%
  \institution{Osaka University}
  \city{Osaka}
  \country{Japan}
}
\email{saru@ist.osaka-u.ac.jp}

\author{Sureel Shah}
\affiliation{%
  \institution{University of California San Diego}
  \city{California}
  \country{United States}
}
\email{sbs001@ucsd.edu}

\author{Dinesh Bharadia}
\affiliation{%
  \institution{University of California San Diego}
  \city{California}
  \country{United States}
}
\email{dineshb@eng.ucsd.edu}

\date{July 2023}

\begin{abstract}
Understanding the location of ultra-wideband (UWB) tag-attached objects and people in the real world is vital to enabling a smooth cyber-physical transition. However, most UWB localization systems today require multiple anchors in the environment, which can be very cumbersome to set up. In this work, we develop \name, providing an accuracy of a few centimeters in many real-world scenarios. This paper will delineate the key ideas which allow us to overcome the fundamental restrictions that plague a single anchor point from localization of a device to within an error of a few centimeters. We deploy a VR chess game using everyday objects as a demo and find that our system achieves $2.4$ cm median accuracy and $5.3$ cm $90^\mathrm{th}$ percentile accuracy in dynamic scenarios, performing at least $8\times$ better than state-of-art localization systems. Additionally, we implement a MAC protocol to furnish these locations for over $10$ tags at update rates of $100$ Hz, with a localization latency of $\sim 1$ ms.
\end{abstract}

\begin{document}

\begin{textblock*}{180mm}(15mm,10mm)
\noindent This paper is accepted by ACM SenSys 2023. The published version is \url{https://doi.org/10.1145/3625687.3625810}. The reference of the paper is  {\it Aditya Arun, Shunsuke Saruwatari, Sureel Shah, Dinesh Bharadia, ``XRLoc: Accurate UWB Localization to Realize XR Deployments,'' Proceedings of ACM Conference on Embedded Networked Sensor Systems (ACM SenSys'23), pp.459--473, 2023.}
\end{textblock*}

\maketitle

\section{Introduction} \label{sec:introduction}

Extended Reality (XR), broadly encompassing virtual, augmented, and mixed reality technologies, can potentially revolutionize fields such as education, healthcare, and gaming~\cite{thomas2012survey,alizadehsalehi2020bim,xi2022challenges}. The primary ethos for XR is to provide immersive, interactive, and realistic experiences for users. A key component of delivering this user experience is to transfer the physical world into the virtual space. For example, our everyday spaces and objects can be transformed into video game assets (like tennis racquets, swords, or chess pieces) for interactive gaming applications.\footnote{\label{fn:demo} in a demo, we transform mugs and a desk into our lab to a life-size chess board (link:\href{https://bit.ly/3q7DKKy}{https://bit.ly/3q7DKKy})} To enable these applications, we find a common thread --- any XR system should localize and track objects in an environment. Specifically, this object-tracking system needs to satisfy three key requirements to realize XR applications:

\begin{figure}[t]
    \centering
        \includegraphics[scale=0.3]{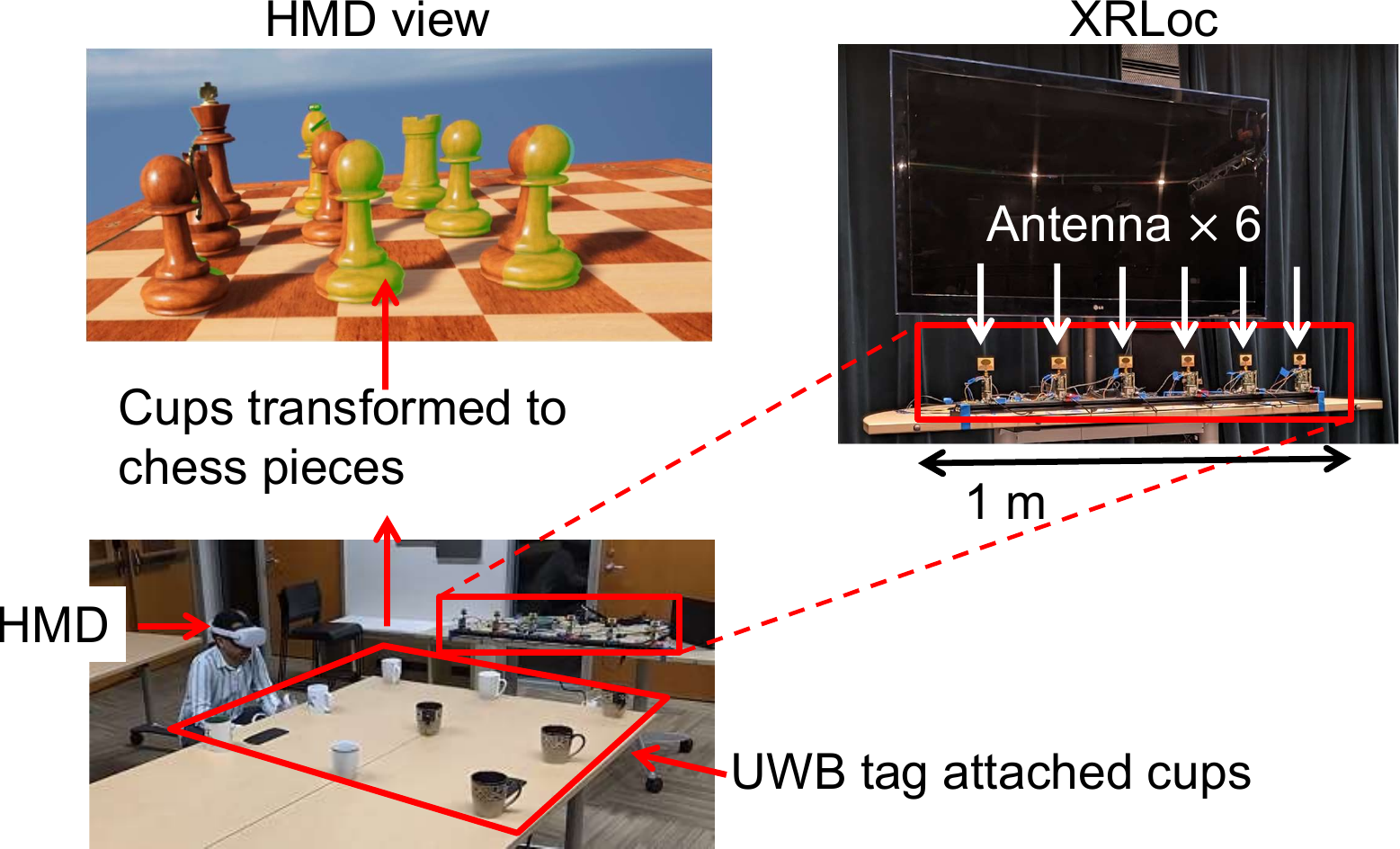}
    \caption{\name enables users to play a life-size chess game with everyday objects. \name localizes mugs retrofitted with off-the-shelf UWB tags from a single vantage point with a few cm of location accuracy, which are then translated to chess pieces in the virtual world.  }
    \label{fig:muloc_prototype_tv}
\end{figure}

\noindent \textbf{R1. Ease of anchor deployment:} Any asset localization system must have low deployment efforts, which can potentially be embedded within common electronics like TVs or soundbars. This single module should be smaller than $1$ m.~\footnote{Most consumer electronics like TVs or soundbars are around $1$ m in length.}  

\noindent \textbf{R2. Accurate and reliable:} Assets must be localized to an accuracy within a few centimeters in room-scale scenarios. We place a stringent requirement of a few centimeters of accuracy to provide a glitch-free user experience. Providing immersive XR experiences consequently means small user or object tracking errors are more obvious and severely impede the adoption of XR~\cite{wang2007design}. Specifically, the localization system must be reliable during movement, under occlusions, and consistently track assets within an accuracy of a few cm.   

\noindent \textbf{R3. Multi-asset low latency localization:} Finally, an XR system needs to localize multiple objects in an environment in real time. In dynamic scenarios, this can mean we must localize tens of objects with a $60$--$80$ Hz update rate as people naturally perceive their surroundings at 60--75 Hz~\cite{deering1998limits}, and delays in updates of object locations in a dynamic scenario can break away from an immersive experience.  

\begin{table*}[]
    \begin{tabular}{|l|c|c|c|c|c|c|c|}
        \hline
         & \textbf{Visual} & \textbf{Acoustic} & \textbf{Radar} & \textbf{RFID} &  \textbf{Single anchor} & \textbf{\name} \\ \hline
        \textbf{R1: Ease of anchor deployment} & \checkmark & \checkmark & \checkmark & $\times$  & \checkmark & \checkmark \\ \hline
        \textbf{R2: Accuracy and reliability} & $\times$ & \checkmark & $\times$ & \checkmark  & $\times$ & \checkmark \\ \hline
        \textbf{R3: Multi-asset and low latency} & \checkmark & $\times$ & \checkmark  & \checkmark & \checkmark & \checkmark \\ \hline
    \end{tabular}
    \caption{Existing technologies do not satisfy the 3 key requirements for an XR localization/tracking system. \label{tab:related-works}}
\end{table*}

However, none of the existing asset localization systems meet these three key requirements to deliver XR applications in everyday scenarios (see Table~\ref{tab:related-works}). 
Camera and visual sensors are susceptible to poor lighting and visual occlusions, consequently failing to provide reliable localization (\textbf{R2}). Additionally, deploying a camera-based system can be privacy invasive~\cite{russell2013people} in home and public settings.
Acoustic systems~\cite{liu2020indoor} provide accurate localization but are difficult to localize multi-asset with low latency simultaneously (\textbf{R3}).
Radar systems~\cite{kong2022m3track,mukherjee2022scalable,xue2021mmmesh} can provide low-latency object tracking from a single module but fail to track occluded objects or those which have small radar cross-sections (RCS). 
Some RFID systems have succeeded in realizing low latency \cite{turbotrack,mobitagbot,tagoram}. Their asymmetric architecture (cost-effective tags and expensive readers) better suits large-scale deployments in retails and industrial sectors. However, long-range RFID systems ($>$ 6m) are expensive and bulky to integrate into consumer electronics, precluding wide-scale deployments (\textbf{R1}).

Altenatively, many single RF module localization solutions~\cite{chen2019m,ge2021single,giorgetti2009single,grosswindhager2018salma,groth2021calibration,kotaru2017position,li2020multipath,meissner2012multipath,soltanaghaei2018multipath,wang2019efficient,wang2019high,zhang2022toward} leveraging WiFi/BLE or ultra-wideband (UWB) are easy to deploy because of transceivers which can be inexpensively deployed in consumer electronics. However, they fail to provide the necessary cm-level accuracy.
None of the existing systems simultaneously satisfy all three stringent requirements to enable XR applications, and prior art will be more carefully considered in Sec.~\ref{sec:related}.

To address the need for XR-compliant localization, we develop \name, which consists of two parts --- a localization tag, attachable to objects of interest, and a single localization module to furnish few-cm level locations from a single vantage point. The localization module is less than $1$ m and can be easily incorporated within everyday electronics such as televisions or soundbars (satisfying \textbf{R1}). It leverages the tag's single UWB transmission for a few cm accurate localization. An accompanying MAC protocol also supports the localization of multiple tags at an update rate of $100$ Hz (satisfying \textbf{R3}). An example deployment of \name is showcased in Fig.~\ref{fig:muloc_prototype_tv}, where beverage cups are attached with off-the-shelf UWB tags. \name is leveraged to transform an office space into a life-sized chess board, with these cups taking the place of chess pieces and localized with cm-level accuracy. A video demo of this case study is also included as well$^{\ref{fn:demo}}$. However, to simultaneously meet all the aforementioned requirements, we need to solve four key challenges:

\noindent \textbf{1. Geometric dilution of precision:} In most UWB localization systems, three or more UWB anchors need to be placed in diverse locations in a room to localize the UWB tag, increasing deployment efforts and breaking away from \textbf{R1}. Alternatively, we can place these UWB anchors within a single localization module constrained to a $1$ m space. However, reducing the spatial diversity can worsen the localization accuracy by $10\times$. This accuracy degradation is called `geometric dilution of precision'~\cite{spilker1996global} (GDOP). A potential strategy to overcome GDOP is to borrow techniques from RFID-systems~\cite{tagoram, turbotrack, mobitagbot} that achieve real-time cm-scale accuracy from a single RFID reader. However, we observe UWB systems provide $15 \times$ worse measurement accuracy compared to RFID systems~\cite{tagoram}, owing to an RFID system sharing the same clock at the transmitter and receiver (mono-static architecture). Hence a direct consequence of GDOP is a \name's reduced resilience to measurement noise, which precludes us from directly borrowing techniques from RFID-based systems.

To reduce our measurement noise, we could increase transmit power to improve signal quality, increase transmission length for better averaging, or choose better hardware with lower noise floors. However, these solutions come at the cost of increased battery consumption at the tag, increased localization latency, or expensive tag design, respectively. Alternatively, \name makes a key observation when looking at the phase difference of the received UWB signal measurements (PDoA) between a pair of anchors ---  PDoA measurement quality can be improved proportionally to the distance between the pair of anchors. This simple observation forms the cornerstone of \name's design and allows us to satisfy the first requirement \textbf{R1}. 

\noindent \textbf{2. Ambiguous location predictions:} However, this improved PDoA measurement quality comes at a detrimental cost --- increasing the anchor spacing creates multiple ambiguous location predictions as phase measurements wrap around at $2\pi$. The changes in these ambiguities mirror the changes in the true location of the tag, and they do not affect tracking systems~\cite{wang2014rf, cao2021itracku}, which leverage phases to provide cm-level tracking accuracy for handwriting recognition. However, incorrectly choosing an ambiguous absolute location can degrade the accuracy by several tens of centimeters and may create glitches within the XR system. 

To predict accurate locations despite phase wrap-around, \name leverages a simple observation --- unlike phase measurements, time of arrival measurements do not suffer from ambiguity. Specifically, the time difference of arrival (TDoA) between a pair of anchors, although inaccurate in furnishing cm-level localization, can help to detect and filter out ambiguities. By cleverly fusing these time-difference and phase-difference measurements, \name can provide cm-level accurate locations from a single UWB transmission and satisfy the second requirement \textbf{R2}. 

\noindent \textbf{3. Measurement bias-aware localization:} However, as we push the envelope on cm-accurate location predictions, we find that hardware biases can corrupt our location estimates and degrade our location accuracy by over $2 \times$. Specifically, through empirical measurements, and as observed in previous studies~\cite{de2019range}, UWB modules~\cite{decawave-dw1000} suffer from a distance-sensitive measurement bias. We model, estimate, and calibrate for these biases via a three-point calibration procedure. We fuse the time and phase measurements with a corrected PDoA and TDoA measurement model by leveraging a particle filter to provide cm-accurate and low-latency location estimates, satisfying \textbf{R2}.

\noindent \textbf{4. High update rate multi-tag operation:} In addition to providing low-latency localization, \name must furnish locations for multiple objects in the environment. Often, the UWB transmissions for localization from multiple tags in an environment can cause packet collisions at \name's module. The collision causes localization failure $25\%$ of the time. We leverage a low-power wireless side channel to alleviate packet collisions to design a power-efficient medium access control (MAC) protocol. Specifically, \name deploys a LoRa-based MAC to support consistent localization for tens of tags at over $80$ Hz, satisfying \textbf{R3}. 

\name brings together these key techniques to build a $1$ m sized module, consisting of $6$ Decawave DW1000~\cite{evb1000} UWB modules for localization, along with a Semtech LoRa SX1272~\cite{sx1272mb2das} to furnish a side-channel for the MAC protocol. Additionally, we prototype a simple UWB + LoRa Tag using the Decawave EVB1000 and a LoRa SX1272. Through extensive evaluations, we find that \name satisfies all the three stringent requirements with
\begin{enumerate}
    \item Static localization error with median and 90th percentile accuracy of $1.5$ cm and $5.5$ cm, an improvement of $9.5 \times$ and $5.2 \times$ from state-of-art systems~\cite{zhao2021uloc}. 
    \item Dynamic localization error with median and 90th percentile accuracy of $2.4$ cm and $5.3$ cm, an improvement of $11 \times$ and $8 \times$ from state-of-art systems~\cite{zhao2021uloc}. 
    \item Localization failure rate of $0.5 \%$ when using the MAC protocol as compared to a failure rate of $~25\%$ without a MAC protocol, a $50\times$ improvement, for $10$ tags operating simultaneously at $100$ Hz location update rate.  
    \item Location compute latency of $1$ ms, allowing for real-time localization ($60$ Hz) of 16 tags. 
\end{enumerate}

\section{Why is this problem hard?} \label{sec:back}

\begin{figure*}[t]
    \centering
    \includegraphics[width=\linewidth]{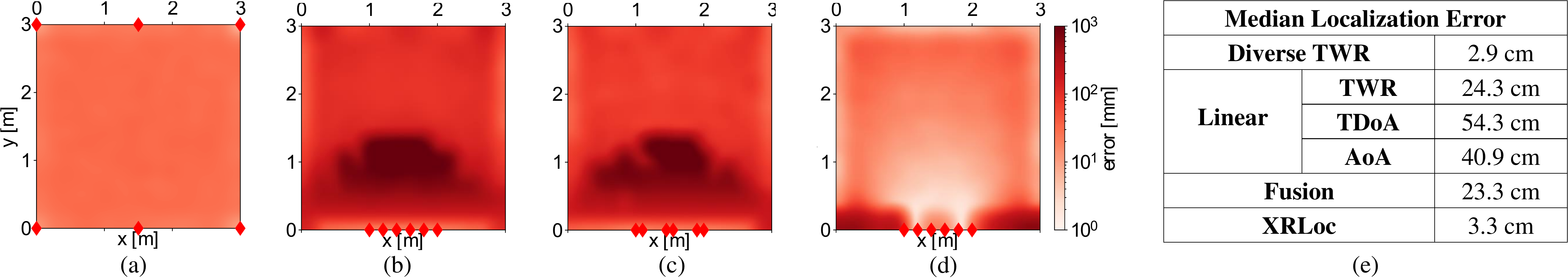}
    \caption{(a) Spatially-diverse placement of UWB anchors (red diamonds) near the walls provides median accuracy with TWR of $2.9$ cm (b) when receivers are constrained near the bottom wall, median accuracy degrades by $~8\times$ when using TWR (c) fusion of TDoA, TWR, and AoA does not help in these scenarios either, providing median accuracy of $23.3$ cm. (d) \name solves the challenges associated with dilution of precision, achieving median accuracy of $3.3$ cm (e) Summary of errors when leveraging various UWB measurements and \name.}
    \label{fig:back}
\end{figure*}

We have established the need for localizing users and objects within a few centimeters of a single vantage point. In this section, we will find that restricting our sensing to within a space of $1$ m reduces our geometric diversity leading to localization errors of many $10$'s of centimeters. This phenomenon is commonly referred to as geometric dilution of precision. We will explore the use of three common UWB measurements -- two-way-ranging (TWR), time-difference-of-arrival (TDoA), and angle-of-arrival (AoA) -- and find systems that rely on these measurements fail to furnish the required accuracy. Additionally, we'll explore fusing and jointly optimizing for these measurements to improve localization accuracy. However, even this measurement fusion is insufficient. To test this hypothesis, we build a simple simulation environment described below. 

\noindent\textbf{Simulation environment:} We perform extensive simulation in a $3 \times 3$ environment, a standard room size, to find the best case localization accuracy. We use $6$ UWB transceivers, placed either diversely in the environment (red diamonds in Fig.~\ref{fig:back} (a)) or in a limited space near the bottom wall (see Fig.~\ref{fig:back} (b, c, d)). Next, we divide this space into a $1$ mm grid and place tags in each position to measure the location accuracy. The pixels of the `heatmaps' represent these tag locations, and the pixel color intensity quantifies the median localization accuracy across $100$ simulated trials.   

\noindent\textbf{Simulating TWR:} Many UWB radios measure the time of flight (ToF) of the signal between the transmitter and receiver with up to a resolution of $15.6$ ps~\cite{decawave-pdoa-kit}. The ToF is measured via multiple packet exchanges, taking at least $0.3$ ms~\cite{corbalan2020ultra}. And clock drifts at the receiver during this TWR event can lead to a ToF measurement deviation of $150$ ps for a $0.5$ ppm clock crystal. Hence, we characterize our simulated TWR measurements with a zero-mean Gaussian with a standard deviation of $150$ ps. 

\noindent\textbf{Simulating TDoA:} Instead of an absolute time of flight measurement, we can measure the difference in the time of arrivals across a pair of synchronized receivers. However, TDoA measurements depend on the receivers' clock synchronization accuracy. Our measurements, independently verified by Decawave~\cite{tdoa_an}, show clock-sync errors in best-case wired synchronization can cause a TDoA measurement deviation of $140$ ps. Hence, our simulated TDoAs are Gaussian distributed with a standard deviation of $140$ ps.  

\noindent\textbf{Simulating AoA:} Some UWB systems~\cite{zhao2021uloc, heydariaan2020anguloc} alternatively measure the angle of arrival of a signal between a pair of receivers placed half-wavelength apart (see close pairs of red diamonds in Fig.~\ref{fig:back}(c)). We can measure AoA with noise deviation of $1.5^\circ$, as independently verified in~\cite{zhao2021uloc, heydariaan2020anguloc}. Consequently, we simulate our AoA measurements as zero-mean Gaussian with $1.5^\circ$ standard deviation.

\subsection{Quantifying localization errors}

TWR, or distance measurements between a tag and multiple receivers placed diversely in an environment, can be used to trilaterate the tag's position to achieve a few cm-level accuracies. From Fig.~\ref{fig:back}(e), we find that the median localization error is $2.9$ cm. Additionally, this error is consistent (with a variation of a few centimeters) across the space (see heatmap in Fig.~\ref{fig:back}(a)). However, when we place all the receivers within a $1$ m linear form factor to satisfy \textbf{R1}, we find that the accuracy degrades by over $8 \times$ as compared to the diverse antenna placement. Additionally, we observe a non-uniform performance with errors as large as $1$ m. To meet \textbf{R2}, we have made our localization system too erroneous to be usable. 

The fundamental reason for the performance degradation is the reduced geometric diversity when the antennas are closer. With the antennas placed around the environment, trilateration is more resilient to errors in distance measurements. We quantify the localization errors by leveraging TDoA or AoA measurements and summarize the results in Fig.~\ref{fig:back}(e). We find under low geometric diversity, the median localization errors can be close to $54.4$ cm and $40.9$ cm for TDoA and AoA measurements, respectively.   

\subsection{Fusing all measurements}

Similar to many robotics applications~\cite{alatise2020review}, we can use TWR, AoA, and TDoA measurements to provide higher accuracy. This fusion is done by jointly optimizing the error function from TWR, AoA, and TDoA~\cite{liu2013joint} measurements. Specifically, in Fig.~\ref{fig:back}(c), we measure $6$ TWR measurements from each receiver (red diamonds), $3$ AoA measurements from each closely-spaced pair of UWB receivers, and $3$ TDoA measurements between one antenna from each of these paired groups. The measurement-fusion efforts provide median localization of $23.3$ cm. However, it still fails to meet our criteria of a few-cm error in localization. 

None of the existing states of art systems can surmount the challenge of localizing from a single vantage point and deliver the stringent requirements set forth by our application use case. In \name, we develop the algorithm (Sec.~\ref{sec:design}) and prototype a system (Sec.~\ref{sec:implementation} and ~\ref{sec:des-mac}) to achieve this small-form-factor, high accuracy (median accuracy of $3.3$ cm as seen from Fig.\ref{fig:back}(d) and Fig.~\ref{fig:back}(e)), and multi-asset localization system, for use within VR systems and immersive audio applications. In the following section, we will delineate the key ideas which allow \name to circumvent the challenges posed by geometric dilution of precision. 

\section{Circumventing low-spatial diversity}
\label{sec:design}

In the following sections, we tackle the fundamental challenge in single-vantage point localization. First, we will explore improving our phase measurements to improve location accuracy by increasing the antenna separation (Sec.~\ref{sec:des-res}). However, this comes with the unintended side-effect of introducing ambiguities to our location prediction. So, we explore the use of time difference of arrival (TDoA) measurements to combat these ambiguities (Sec.~\ref{sec:des-amb}). Finally, we explore fusing these measurements in an accurate and low-latency fashion by leveraging a particle filter (Sec.~\ref{sec:des-opt}). By exploring the key ideas here, \name will fulfill \textbf{R2} and furnish few-cm level localization.

\subsection{Improving localization resolution}\label{sec:des-res}

The prior learning from Sec.~\ref{sec:back} is that we reduce our resiliency to noise when we try to localize tags from a single vantage point.
Lacking spatial diversity adds vulnerability to the optimization creating large outlier measurements and preventing few-cm scale localization. However, when we have two closely (less than half-wavelength) separated antennas, we can find the phase difference ($\Delta \phi$) between this pair as 
\begin{align*}
\Delta \phi = \frac{2\pi d}{\lambda} \sin(\theta)
\end{align*}
where $\theta$ is the incoming angle of arrival w.r.t. to the normal of this pair of antennas, $d = \frac{\lambda}{2}$ is the distance between them, and $\lambda$ is the wavelength at $3.5$ GHz UWB center frequency.\footnote{we develop this intuition assuming far field, but later in Sec.~\ref{sec:des-opt} we consider the exact phase difference measurement} However, the typical UWB phase has a resolution of around $8$ bits \footnote{For example, although having a 16-bit real part and a 16-bit imaginary part in each CIR sample, DW1000 has a 7-bit phase resolution because the phase must be corrected by RCPHASE register, which is a 7-bits and for the adjustment of receiver carrier phase.}, which provides a phase resolution of $1.4^\circ$, and consequently a localization resolution of $~2.1$ cm at a distance of $3$ m from the localization module. 
However, increasing the inter-antenna separation, $d$, linearly increases the measured phase difference. We can leverage this to improve our localization resolution to the $\sim 1$ mm limit when the antenna separation is $1$ m.

Prior works~\cite{wang2014rf, cao2021itracku} have leveraged this fact to increase accuracy for handwriting tracking purposes. But, widening this separation comes at the cost of introducing more phase ambiguities. This is apparent when we return to the AoA equation and observe that our phase-difference measurements, $\Delta \phi$, wrap over $2 \pi$ for a larger separation than half-wavelength separation for angles between $-90^\circ < \theta < 90^\circ$. This is not an issue for tracking purposes, where the changes in location of these ambiguities mirror the true changes in the location and continue to provide a similar trajectory estimate. However, for \name, we find predicting and tracking these incorrect locations can degrade the localization accuracy by an order of magnitude to several tens of centimeters.

\subsection{Ruling out ambiguities} \label{sec:des-amb}

\begin{figure}[t]
    \centering
    \includegraphics[width=\linewidth]{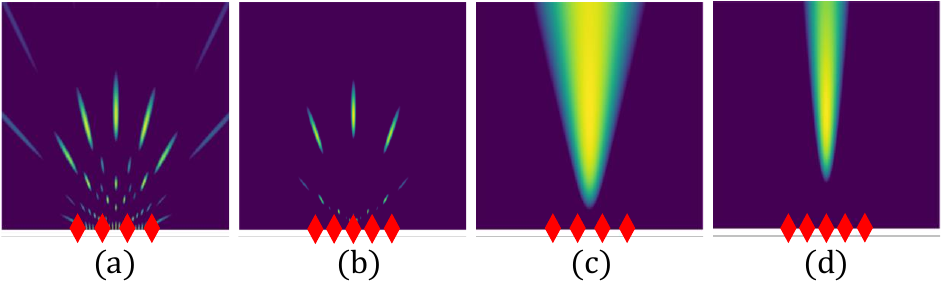}
\caption{Log-likelihood heat map of PDoA and TDoA when changing the number of antennas $N$.}
\label{fig:pdoa-tdoa-change-N}
\end{figure}

To overcome ambiguities, a simple solution is adding more antennas between the two we have placed so far. These additional antennas will help eliminate phase ambiguities by reducing the consecutive antenna distance while employing a 1-m antenna array aperture. Fig.~\ref{fig:pdoa-tdoa-change-N} (a, b) depicts these ambiguities that exist in such a system by showing the likely positions of the tag. Considering the simulation environment from Sec.~\ref{sec:back}, we deploy two arrays with spacing $33.3$ cm and $25$ cm for the same antenna aperture of $1$ m. Next, we deploy a tag at the center of the space and predict its potential locations (pixel color intensities) in both scenarios. We observe that keeping the same aperture of 1 m, we have similar measurement errors (peak widths) in both cases, consistent with our previous findings, but reducing separation creates fewer ambiguities. Deploying $23$ antennas within this $1$ m, each spaced half-wavelength apart, will remove all our ambiguities at the cost of increased hardware complexity.  

Alternatively, we observe TDoA measurements are free from ambiguities and can potentially be leveraged to disambiguate the predictions from PDoA. Similarly to the previous PDoA images, in Fig.~\ref{fig:pdoa-tdoa-change-N} (c, d), we only show the tag's location likelihoods when relying on TDoA measurements. The TDoA peak, although very erroneous (larger peak widths), is unambiguous. Additionally, increasing the number of antennas reduces this error/peak width. To recap, by reducing the antenna separation (or increasing the number of antennas), we increase the separations between the ambiguities coming from PDoA measurements and tighten our peak widths coming from TDoA. Consequently, at the correct antenna spacing, our ambiguous peaks will be wide enough to be rejected by our TDoA measurements. We find this sweet spot when we use $6$ antennas, $4\times$ fewer antennas than would have otherwise been required. 

\subsection{Jointly optimizing for TDoA and PDoA measurements}\label{sec:des-opt}

We can now extend the key intuitions to leverage TDoA and PDoA to develop a localization algorithm to meet our few-cm accuracy requirement. As further explained in Sec.~\ref{sec:implementation}, via careful engineering and hardware design choices, we measure PDoA with a standard deviation $\sigma_\theta = 5^\circ$ and TDoA with a standard deviation of $\sigma_t = 150 ps$. This \textit{measurements} can be modeled as a zero-mean Gaussian:  
\begin{align*}
    \mathrm{TDoA\ between\ Rx\ i\ and\ j: }\ &t_{i,j} \sim \mathcal{N}(0, \sigma_t) \\
    \mathrm{PDoA\ between\ Rx\ i\ and\ j: }\ &\theta_{i,j} \sim \mathcal{N}(0, \sigma_\theta).
\end{align*}
Additionally, given a candidate tag location, $\vec{p}$, and receiver locations $\vec{x_i}, \forall i \in [1, 2,  \dots, N]$ we can also compute the \textit{expected} PDoA and TDoA as 
\begin{align}
    \mathrm{TDoA:}\ &\hat{t}_{ij} = \frac{|\vec{p} - \vec{x_i|}}{c} - \frac{|\vec{p} - \vec{x_j}|}{c} \nonumber \\ 
    \mathrm{PDoA:}\ &\hat{\theta}_{ij} = \mod \left(2\pi \left(\frac{|\vec{p} - \vec{x_i}|}{\lambda} - \frac{|\vec{p} - \vec{x_j}|}{\lambda}\right), 2\pi\right) \label{eq:pdoa}
\end{align}
where $\vec{p}$ is the location of the tag and $\vec{x}_i / \vec{x}_j$ are the locations of the 6 UWB antennas placed within a linear $1$ m array. $c$ and $\lambda$ are the speed of light and UWB wavelength, respectively. Note here we forgo the far-field assumption made in Sec.~\ref{sec:des-res}. 

The location ($\vec{p}$) which gives the closest expected measurements to the actual measurements is the likely tag location, 
\begin{align}\label{eq:opt}
    \min_{\vec{p}} \begin{bmatrix} \vec{e_t}^T & \vec{e_\theta}^T \end{bmatrix}^T \Sigma^{-1}\begin{bmatrix} \vec{e_t}^T & \vec{e_\theta}^T \end{bmatrix} 
\end{align}
where $\vec{e}_t$ and $\vec{e}_\theta$ measure the error between our predictions and the actual measurements, and 
\begin{align*}    
\Sigma = \mathrm{diag}(\sigma_t^2, \cdots, \sigma_t^2, \sigma_\theta^2, \cdots \sigma_\theta^2)
\end{align*}
is a diagonal covariance matrix containing the TDoA and PDoA measurements standard deviations. Note here that since each receiver on \name's localization module is independently measuring the TDoA and PDoA, we have a diagonal covariance matrix. 

The simplest way to find this best tag location is to perform a grid search over our space to find the minimum point for Eq.~\ref{eq:opt}. Aiming for cm-level localization, we choose a grid size of $1 \times 1$ mm. But this exhaustive search can be time-consuming (around $61.2$ s / location on a 12-core CPU), precluding real-time localization in dynamic situations. Alternatively, we can leverage gradient descent-based optimization techniques~\cite{gil2007numerical} to arrive at the most likely tag position. However, these techniques fail when we do not have a good initial estimate of the location, which is the case when looking to localize a tag in a large environment~\cite{glorot2010understanding}.  

To surmount this challenge, we provide the final insight ---selectively searching over the large space instead can reduce the computation complexity for localization. The brute force approach unnecessarily searches over each grid point for every packet. We can instead sample our environment more sparsely and slowly converge to our ideal location over a few packets. This is, in fact, the key idea behind particle filters~\cite{aernouts2020combining}, which are commonly used in state estimation scenarios with highly non-convex error functions and poor initialization. 

Armed with this insight, for the first packet we receive, we uniformly distribute a set of particles (500 particles/$\mathrm{m}^2$) in our environment and compute the likelihood of these positions. When we receive consecutive packets, we can re-sample the set of particles with the highest probability and continue converging to our true locations. However, despite the fewer likelihood computations required, particle filters commonly furnish non-real-time estimates (with a latency of $~7.2$ ms on a 12-core CPU). To combat this problem, \name adaptively re-samples and reduces the number of particles based on the current confidence of the estimate. As we do not know the tag's location, many particles are initially required to sample the search space uniformly. However, our particles converge close to the true location over time, improving our confidence in the location estimate. We can reduce the number of particles needed as we no longer need to explore the space uniformly. Empirically, this adaptive particle filter implementation converges within five measurements and provides a location estimate with a $1.2$ ms latency on a 12-core CPU.

\section{Challenges with prototyping \name} \label{sec:implementation}

Additional considerations arise when employing the ideas from Sec.~\ref{sec:design} while prototyping \name using off-the-shelf components. First, we need to acquire low-noise phase measurements. In Sec.~\ref{im:clock}, selecting the right clock is imperative to ensure a low phase noise. Second, due to hardware imperfections, we find that the expected PDoA measurements (Eq.~\ref{eq:pdoa}) do not match the real-world measurements. To account for the offsets, we devise a calibration scheme and re-consider the formulation of the expected PDoA measurements in Sec.~\ref{im:calib}. Finally, we explore the effects of multipath reflection on the TDoA measurements in Sec.~\ref{im:mp}.

\subsection{Acquiring accurate time and phase} \label{im:clock}

\begin{figure}
    \centering
    \includegraphics[width=\linewidth]{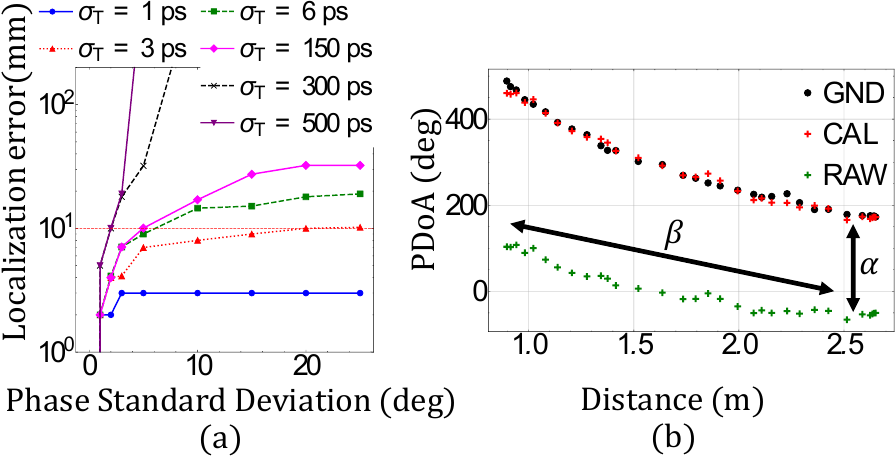}
    \caption{ (a) Localization error vs. PDoA error standard deviation, with TDoA error standard deviations as each line in the legend. For few-cm level localization, the threshold, per the red line, is $\sigma_t = 5^\circ$ and $\sigma_t = 150$ ps. (b) Phase measurements (green) deviate from ideal (black) measurements. Performing appropriate calibration fixes these deviations (red).}
    \label{fig:error_param_study}
\end{figure}

Before prototyping \name, we conducted extensive simulations to investigate the minimum phase and time acquisition accuracy needed to achieve few-centimeter positioning accuracy, assuming 6 antennas were equally spaced in a 1-meter region. In a $3 \times 3$ environment, we implemented the algorithm presented in Sec.~\ref{sec:des-opt} at varying phase and time acquisition noise levels. 
Our simulation results are presented in Fig.~\ref{fig:error_param_study}(a), where the horizontal axis represents the standard deviation of the phase error, and the vertical axis represents the 50 percentile of the localization error. Each line shows the standard deviation of the time error.

From this simulation, we make two key observations. First, we see that time errors between 3--250 ps provide similar localization accuracy, and these lines are grouped in the plot. However, exceeding $300$ ps in time error significantly increases localization error, as TDoA fails to segregate ambiguity made by PDoA. Second, these simulations clarify that few-cm level accurate localization requires high phase accuracy. Specifically, the red vertical line marks a threshold of $5^\circ$ of standard deviation in phase measurement needed to achieve few-cm accurate locations. 

The synchronization clock is the main factor affecting this phase noise in our system. The phase of the UWB signal is measured by first down-converting the received signal with the carrier signal. It is measured relative to this carrier signal by the baseband processing unit~\cite{decawave-pdoa-kit}. And when we consider measuring the PDoA, we look at the difference in the phase of any two receivers. In this situation, if both receivers share the same carrier clock, then the PDoA they measure will be induced purely from the relative distance traveled by the signals to each receiver (see Eq.~\ref{eq:pdoa}). A simple way to achieve this is to connect the two receive antennas to the same UWB module~\cite{cao2021itracku}. However, we observed the overhead of extracting the complete CIR when implementing these systems is large ($\sim 1.2$ ms), precluding low-latency localization. Specifically, we have the API overhead to measure the data and the data extraction overhead over USB, requiring 599 $\mu$s and 612 $\mu$s, respectively. 

Alternatively, we prototype our system using independent UWB modules~\cite{evb1000} for each receiver, eliminating the need to export CIR measurements. This reduces the data acquisition latency by $\sim 4 \times$ to $\sim 340~\mu$s. However, we cannot synchronize the carrier clocks on these independent modules, but instead, synchronize a lower $38.4$ MHz clock leading to phase measurement errors. Via measurements with different clocks, we find that the phase noise in this input clock can largely influence the noise in the PDoA measurements. Specifically, from the oscillator's data sheet~\cite{crystek}, we can obtain the phase noise of the oscillator, $N_\phi (f_{\rm offset})$ where $f_{\rm offset}$ is the frequency offset from the center frequency of the oscillator.
Using the $N_\phi (f_{\rm offset})$, the standard deviation of clock jitter, $\sigma_{\rm jitter}$, can be expressed as follows.
\begin{align}
    \sigma_{\rm jitter} =  \frac{\sqrt{2}}{2 \pi f_{\rm osc}} \sqrt{\Delta f N_\phi (f_{\rm offset})}
\end{align}
where, $\Delta f$ is the bandwidth of the measurement and $f_{\rm osc}$ is the oscillator frequency. We measure the standard deviation of the phase error ($\sigma_\phi$) and time stamping error ($\sigma_t$) as:
\begin{align*}
    \sigma_{\phi} = \frac{c}{\lambda} \frac{f_{\rm osc}}{2 \pi f_s} \sigma_{\rm jitter}\ ; \quad \sigma_{t} = \frac{f_{\rm osc}}{f_{t}} \sigma_{\rm jitter}
\end{align*}
where, $f_s$ is the sampling frequency, $f_{\rm t}$ is the frequency of the clock used for to measure time-of-arrival and $c$ is the speed of light. We can choose an appropriate clock to meet our phase and time measurement thresholds by modeling this noise behavior. Many off-the-shelf~\cite{crystek, astxr} clocks satisfy these requirements at reasonable price points and employ \cite{crystek} in prototyping \name.
For example, according to the datasheets provided by Crystek \cite{crystek} and Abracon \cite{astxr}, their respective phase noise values at 100 kHz offset are $-160$ dBc/Hz and $-150$ dBc/Hz, while their respective phase noise values at 100 Hz offset are $-115$ dBc/Hz and $-109$ dBc/Hz.

\subsection{Combating hardware biases}\label{im:calib}

In Eq.~\ref{eq:pdoa}, we provided an expression for the expected PDoA measurement if we know the underlying tag and receiver locations. In reality, however, we see a large deviation when we compare the expected PDoA measurements with true PDoA measurements. To verify this, we perform an experiment varying the distance of a tag from \name's localization module. In Fig.~\ref{fig:error_param_study}(b), the green`RAW' measurements are shifted from black ground truth `GND' measurements. Visually, we observe three deviations --- a constant additive bias ($\alpha$) which contributes to a downward shift, a multiplicative bias ($\beta$) w.r.t. distance affecting the slope of the line, and an exponential bias ($\gamma$) w.r.t. distance affecting the curvature (non-visualized in the figure). We assume these biases result from the ADC saturation when the distances are too close and propose a 3-point calibration to compute these hardware-specific calibrations below. Subsequently, we modify our \textit{expected} PDoA measurements from Eq.~\ref{eq:pdoa} as 
\begin{align*}
    \hat{\phi}_{i,j} = \mod\bigg( & \left\{\frac{2\pi d_i}{\lambda} - \alpha_i - \beta_i d_i^{\gamma_i}\right\} -  \\
                                    & \left\{\frac{2\pi d_j}{\lambda} - \alpha_j - \beta_j d_j^{\gamma_j}\right\}, 2\pi\bigg)
\end{align*}
where, $\alpha_i$, $\beta_i$, $\gamma_i$ are the calibration parameters and $d_i = |\vec{p} - \vec{x_i}|$ is the distance between the tag and UWB receiver. 
We replace Eq.~\ref{eq:pdoa} with this updated expected PDoA equation for the particle filter described in Sec.~\ref{sec:des-opt}. 

To estimate these calibration parameters, we perform a three-point calibration. First, we model the phase ($\tilde{\Phi}$) measured at each UWB module according to these biases as
\begin{align*}
\tilde{\Phi}_i = \Phi_{i} + \alpha_i + \beta_i (d_i)^{\gamma_i}, \quad i \in [1, N],
\end{align*}
where $\tilde \Phi$ is the calibrated phase. Next, we measure the received phase ($\Phi$) at each UWB receiver for three \textit{known} locations within our space. Finally, we use regression to find the expected calibration parameters, which minimize the deviation between the measured and expected phases according to the above equation. 

\subsection{Handling multipath reflections} \label{im:mp}

However, in common indoor settings, reflections of the RF signal can potentially lead to ambiguities in TDoA measurement~\cite{scheuing2006disambiguation}. Despite our best efforts to acquire bias-corrected PDoA measurements, the presence of multipath can prevent us from ruling out ambiguous location predictions. However, UWB signals sample at the rate of $1$ GHz, implying a time resolution of $1$ ns. This fine-time resolution implies we are only corrupted by reflected paths whose additional travel distance is within $30$ cm. In indoor environments, finding such close-by reflected paths is unlikely, and we find that our direct path and reflected signals are separable in the time domain. With this in mind, we measure the time of arrival and phase of the signals at the hardware reported first peak index, FPI~\cite{dw_fpi}, at the 6 UWB receivers in \name's localization module.   

\begin{figure*}[t]
    \centering
    \includegraphics[width=1.00\linewidth]{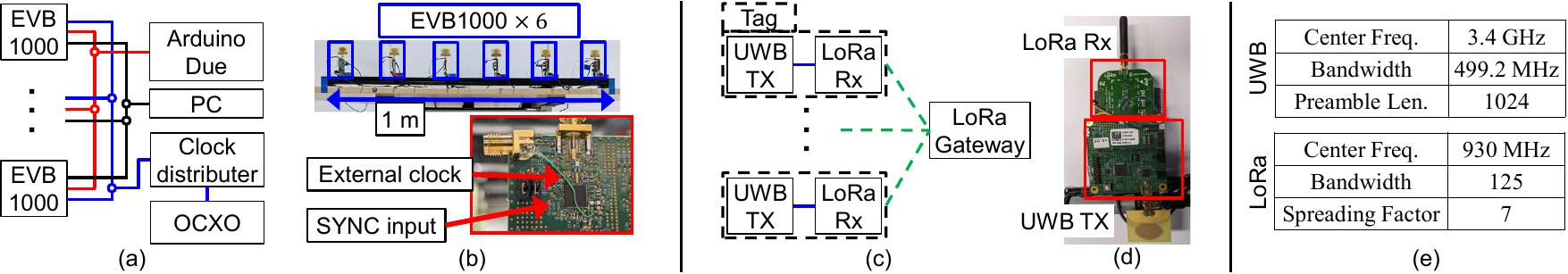}
    \caption{Implementation: \textbf{(a)} Block diagram showcasing interconnections between the $6$ UWB receivers~\cite{evb1000}, the clock synchronization scheme (blue), ``SYNC'' implementation (red), and data back-haul via USB (black). \textbf{(b)} real-world implementation of block diagram; \textit{inset}: external modification to UWB receiver. \textbf{(c)} block diagram for Tag showcasing the UWB and LoRa radios, the interrupt line (blue) to schedule UWB transmission and LoRa clock-sync broadcasts (dotted green) (d) real-world implementation of Tag. (e) UWB/LoRa radio parameters.}
    \label{fig:implementation}
\end{figure*}

\section{Enabling multi-tag operation} 
\label{sec:des-mac}

Through the ideas presented in Sec.~\ref{sec:design} and ~\ref{sec:implementation}, \name fulfills the first two requirements for a localization system to be compatible with XR applications --- ease of deployment (\textbf{R1}) and accuracy (\textbf{R2}). However, when we extend the current system to localize multiple tags in an environment, packet collisions amongst various tags can detrimentally affect our localization rates, resulting in a packet drop of $25\%$. Alternative to allowing tags to transmit arbitrarily, we can schedule individual tags at specific time intervals and leverage time-division multiple access (TDMA) to prevent collisions. 

We seek to enable a total localization rate of $1000$ Hz at \name's receiver means localizing a $1000$ tags at a rate of $1$ Hz or $10$ tags at $100$ Hz.
Specifically, we explore leveraging low-power wireless technologies~\cite{sanchez2016state,gupta2016ble} as a side channel for MAC protocol operation. 
A MAC controller needs to perform three tasks --- onboarding new tags, providing time synchronization, and applying corrections to tags that deviate from their time slots. Existing systems~\cite{tiemann2019atlas, macoir2018mac, bauwens2021uwb} leverage UWB signals for providing this MAC control. However, we observe when a large number of tags need to be onboarded or corrections to the tag's time slots need to be made, frequent collisions between UWB beacons for localization and UWB transmission for MAC control can exacerbate the problem we seek to solve. Alternatively, we propose using an additional side-channel leveraging low-power wireless technologies~\cite{sanchez2016state,gupta2016ble} to simplify the MAC control and allow for independent tag management and localization functions.
UWB is known to have high power consumption (e.g., about 416 mW for DW1000) during reception due to its use of wide bandwidth and despreading processing \cite{biri2020socitrack}.
From the viewpoint of low power reception, using LoRa (e.g., about 20 mW for SX1280) or BLE (e.g., about 16 mW for nRF52832) for the side channel is practical.
We employ LoRa as a side channel to furnish reliable and low-power MAC control for multiple tags.
LoRa and UWB are at $~900$ MHz and $~3.5$ GHz, allowing them to co-exist with minimal interference. We also note that alternative side channels like BLE be employed; however, we choose to implement this prototype with LoRa given its simplicity. 

The MAC protocol consists of two components --- a LoRa MAC controller (gateway), which is deployed along with the localization module we have built so far, and a LoRa Receiver (LoRa RX) connected to the UWB tag. The gateway performs the three core functions of the MAC protocol. 

\noindent \textbf{Discovery and Onboarding:} New tags introduced to a system transmit beacon packets to announce their presence. Subsequently, the gateway invites these new tags to join the network by assigning a specific transmit time slot to transmit the UWB localization packets. The number and duration of a transmit slot are determined by the maximum number of tags and their localization rate. Currently, we support $1000$ slot with a $1$ ms slot width. Fig.~\ref{fig:implementation}(c) illustrates a block diagram of operation. 

\noindent \textbf{Global Time sync:} Each tag must have a consistent notion of time slots, which requires a global time synchronization within the accuracy of at least half the slot width. Previous works~\cite{ramirez2019longshot} have showcased $\mu$s accuracy in synchronization clocks, and we leverage these works to provide time synchronization. Specifically, the gateway transmits time-sync packets every $100$ s, the time it takes for the $5$ ppm clocks to drift by $500$ $\mu$s, half the slot width for each tag. The LoRa RX receives these sync packets and corrects for its clock drift. 

\noindent \textbf{Correcting erroneous tags:} Finally, as a precautionary measure, we develop a correction mechanism to re-slot colliding tags. There may be a time-sync failure at tags, resulting in transmission at an incorrect time slot, leading to consistent collisions among groups of tags. By tracking the tags which suffer consistent collisions, the gateway broadcasts a correction packet over LoRa to re-slot the erroneous tag.   

\section{Implementation}

We have seen \name consists of three core components --- the localization module, the LoRa MAC handler, and the UWB+LoRa tag. This section will take a closer look at prototyping these components. 

\noindent \textbf{Localization Module:} \name's primary contribution is a single-vantage point localization module using off-the-shelf components with a size of $1$ m. This small size allows the localization module to be deployed within common electronics like TV's our soundbars.  
Fig.\ref{fig:implementation} shows the implemented prototype.
The prototype is built with $6$ UWB receivers EVB1000~\cite{evb1000}, with table~\ref{fig:implementation}(e) detailing the configuration parameters. We synchronize the UWB modules to a common clock (OCXO~\cite{crystek}) via a clock distributor module~\cite{lmk-clock} as shown by the `blue' path in Fig/~\ref{fig:implementation}(a).  
Additional to the clock modification discussed in Sec.~\ref{im:clock}, we expose the EVB1000's `SYNC' pin to reset the time on the UWB modules to reduce bias in TDoA measurements. This sync is handled by an Arduino Due and is indicated in the `red' path, with additional details provided in \cite{tdoa_an}. 
When each EVB1000 receives a single ``blink'' signal for localization from the UWB Tag, the receiver reports the first-peak-index (FPI) of the direct path in the channel impulse response's peak, the signal phase at this point, time of arrival (RXTIME), and a carrier phase correction (RCPHASE) via the data path (shown in black).

\noindent \textbf{LoRa MAC gateway:} The LoRa gateway is the central controller to initialize, discover, and onboard all the tags in the environment. It is prototyped with a LoRa SX1272~\cite{sx1272mb2das} transmitter. This handler maintains the MAC state machine and performs all the functions described in Sec.~\ref{sec:des-mac}.  

\noindent \textbf{Tags:} We prototype the tag (shown in Fig.~\ref{fig:implementation}(d)) using the EVB1000~\cite{evb1000} and program it with the parameters in Table~\ref{fig:implementation}(e). 
The tag transmits `blink' packets at $60$ Hz, with each transmitted frame having 14 bytes of payload, including packet number and MAC address, to facilitate and test the MAC protocol. Operating in parallel, we have the LoRa SX1272 receiving time-sync packets from the Gateway module maintaining the UWB transmit slots and providing medium access control. An interrupt pin is raised by LoRa RX (shown in blue in Fig.~\ref{fig:implementation}(d)) to initiate a UWB `blink' transmission at the accurate time slot.

\section{Evaluation}

\begin{figure}
\includegraphics[scale=0.32]{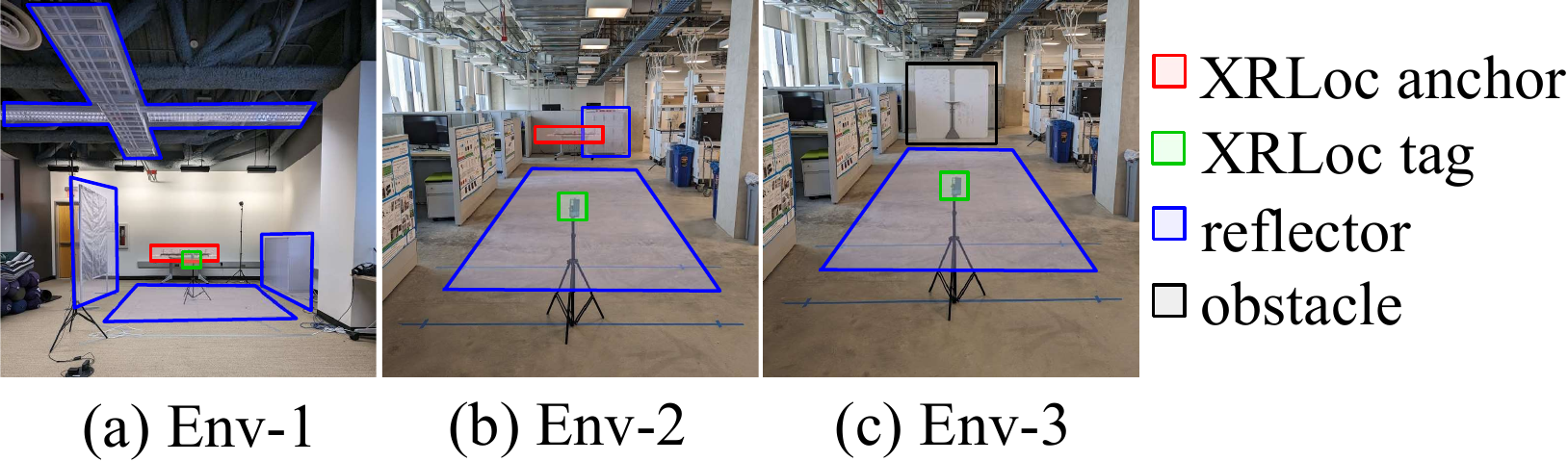}
 \caption{Evaluated in three spaces referred to as (a) Env-1: office-like, (b) Env-2: large-scale, and (c) Env-3: Non-line-of-sight condition. The tag, \name's module, and relevant regions are marked.}
\label{fig:eval-env}
\end{figure}

\name takes strides in achieving a few cm-scale localization in static and dynamic conditions. 
We rigorously test the system over eight different moving datasets and at multiple static points in various environments, including line-of-sight (LOS at Env-1 and 2) and non-line-of-sight (NLOS at Env-3) conditions as shown in Fig.~\ref{fig:eval-env}. 
To make the NLOS condition in Env-3, a wooden board 2.5 cm thick was placed 30 cm forward from the XRLoc anchor.
Additionally, we re-implement state-of-art AoA-based UWB localization system ULoc~\cite{zhao2021uloc} based on their open-source documentation. We place $3$ anchors in a diverse scenario, as a triangle in this space, and a constrained linear scenario, in a $1$ m straight line. We test ULoc with the same static and dynamic positions.  

\begin{figure}
    \newcommand{\baseevalcale}{0.16}

    \begin{subfigure}{0.5\linewidth}
            \includegraphics[scale=\baseevalcale]{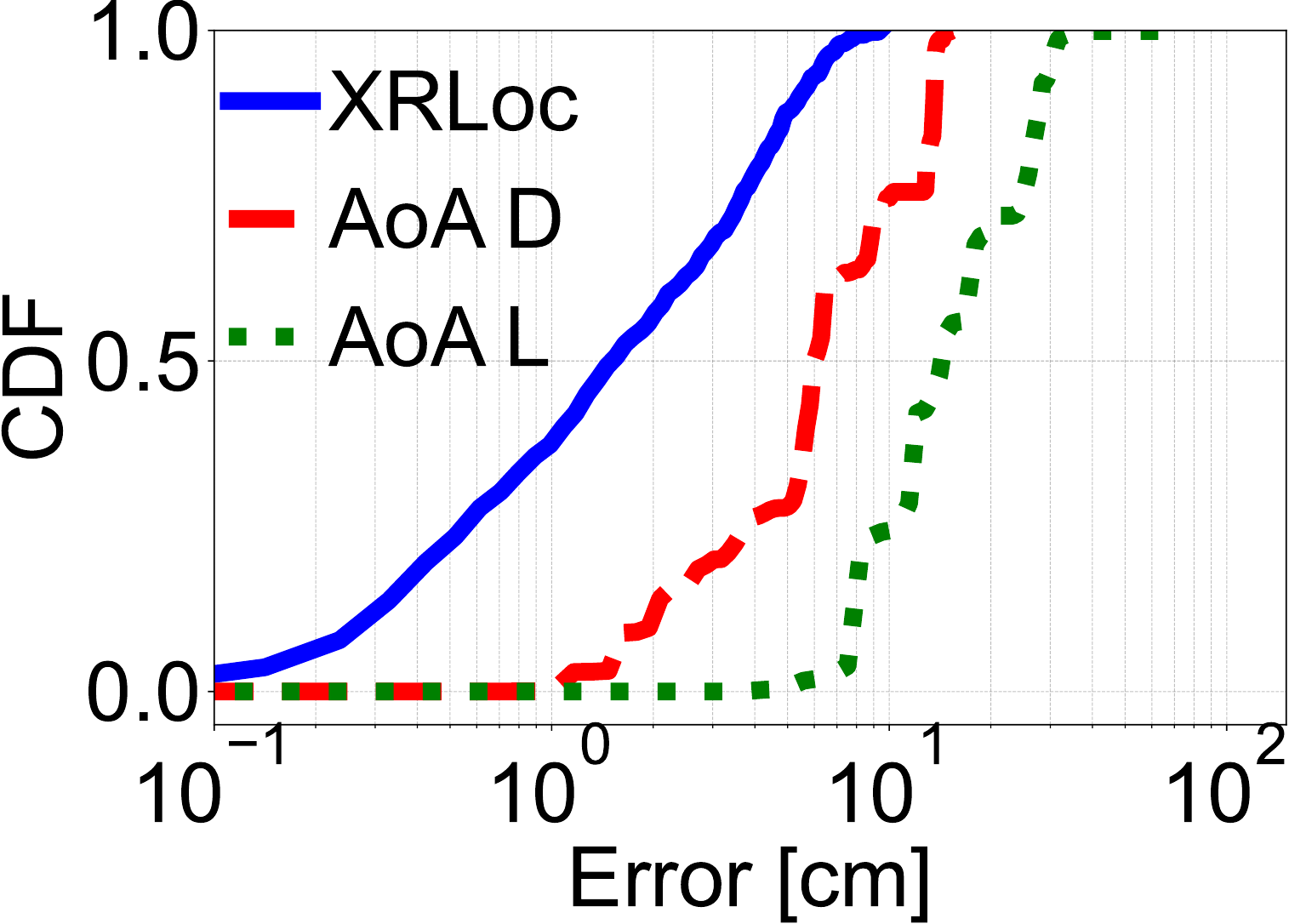}
            \caption{}
    \end{subfigure}
    ~
        \begin{subfigure}{0.5\linewidth}
            \includegraphics[scale=\baseevalcale]{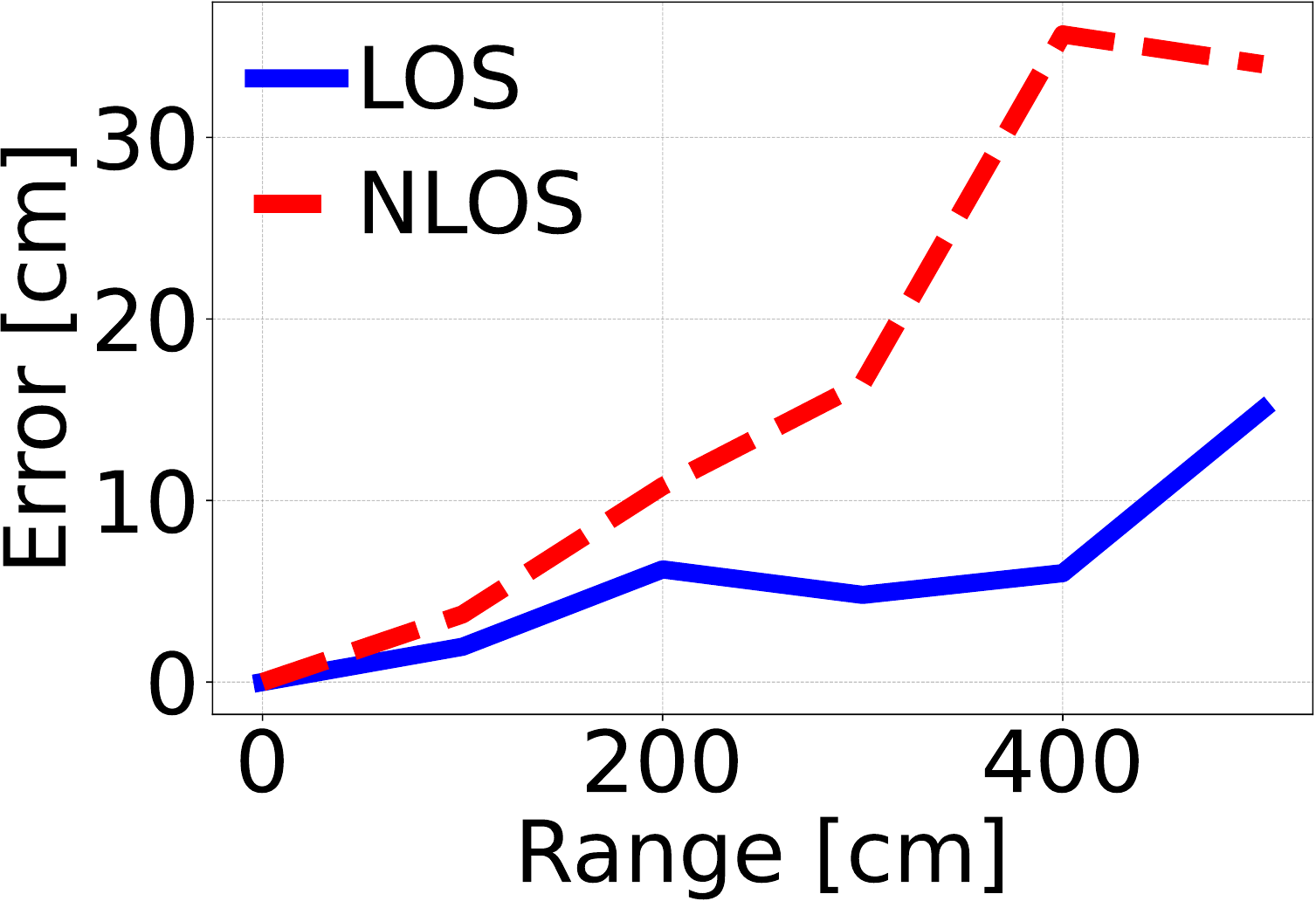}
            \caption{}
    \end{subfigure}
    \caption{\name's Localization performance. (a) Static localization error, (b) localization accuracy vs. range with LOS (Env-2) and NLOS (Env-3) conditions.}
    \label{fig:eval-static}
\end{figure}

\begin{figure*}[t]
    \newcommand{\dynamicevalcale}{0.17}
    \centering
    \begin{subfigure}{0.2\linewidth}
            \includegraphics[scale=\dynamicevalcale]{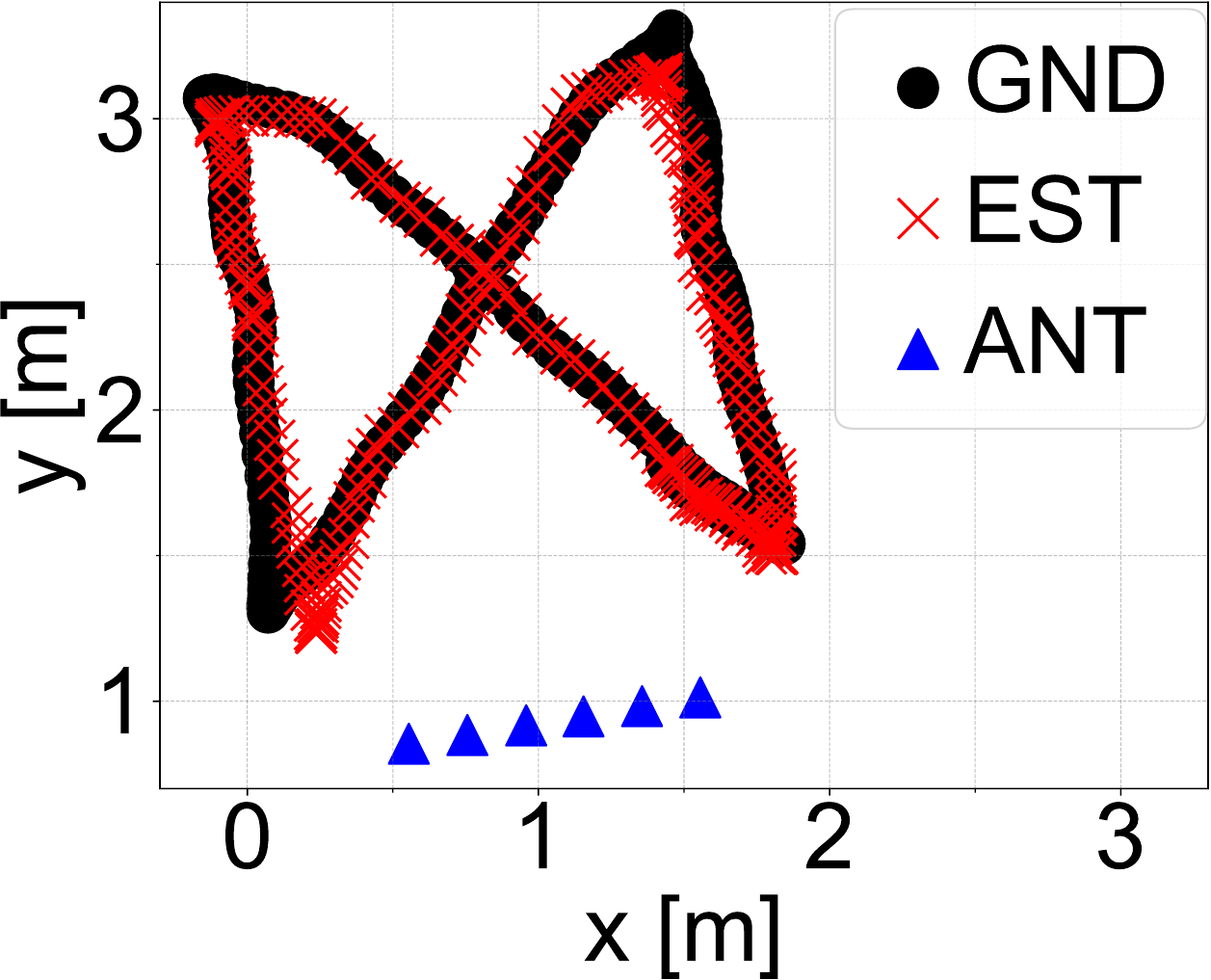}
            \caption{}
    \end{subfigure}
    ~
    \begin{subfigure}{0.2\linewidth}
            \includegraphics[scale=\dynamicevalcale]{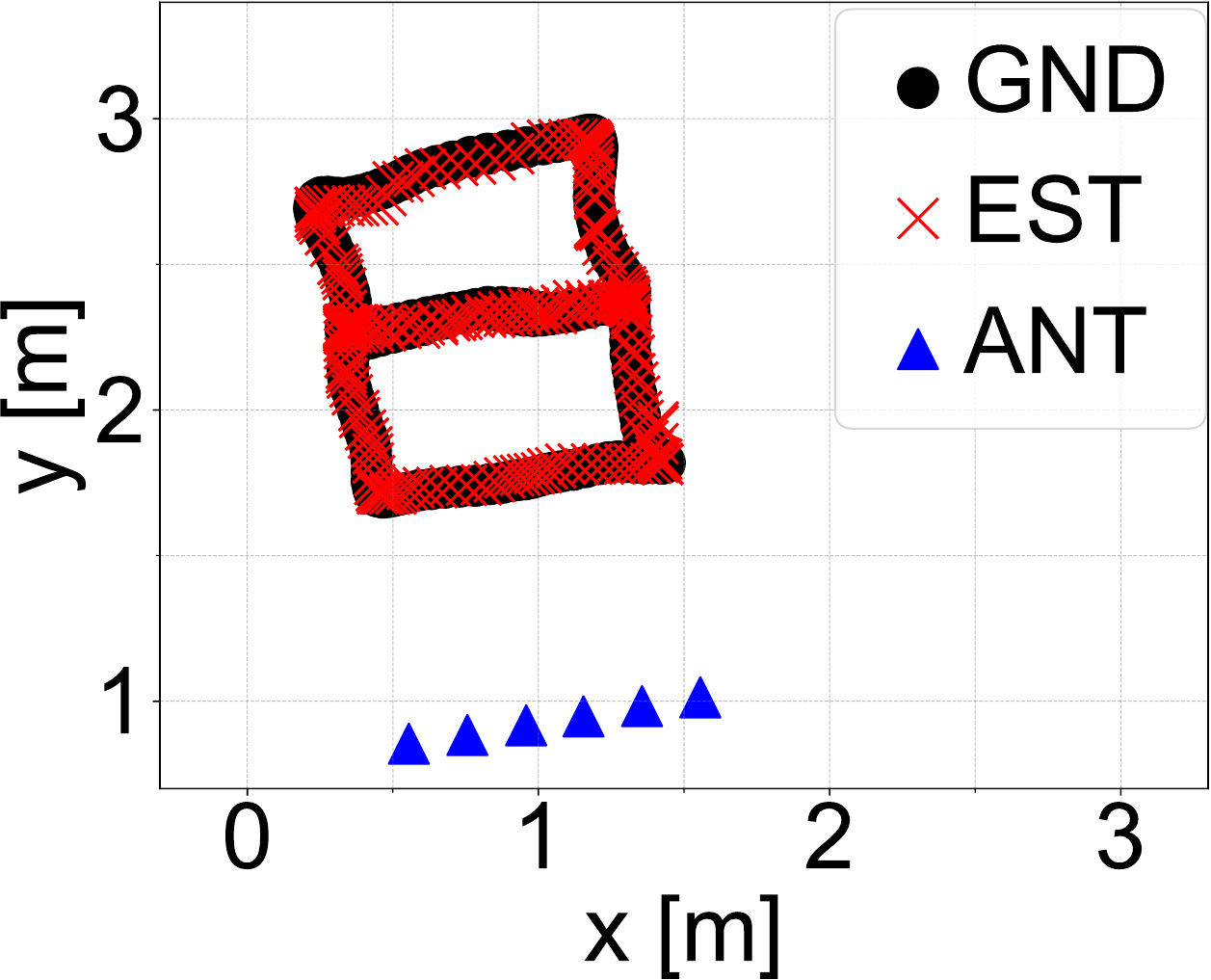}
            \caption{}
    \end{subfigure}
    ~
    \begin{subfigure}{0.26\linewidth}
            \includegraphics[scale=\dynamicevalcale]{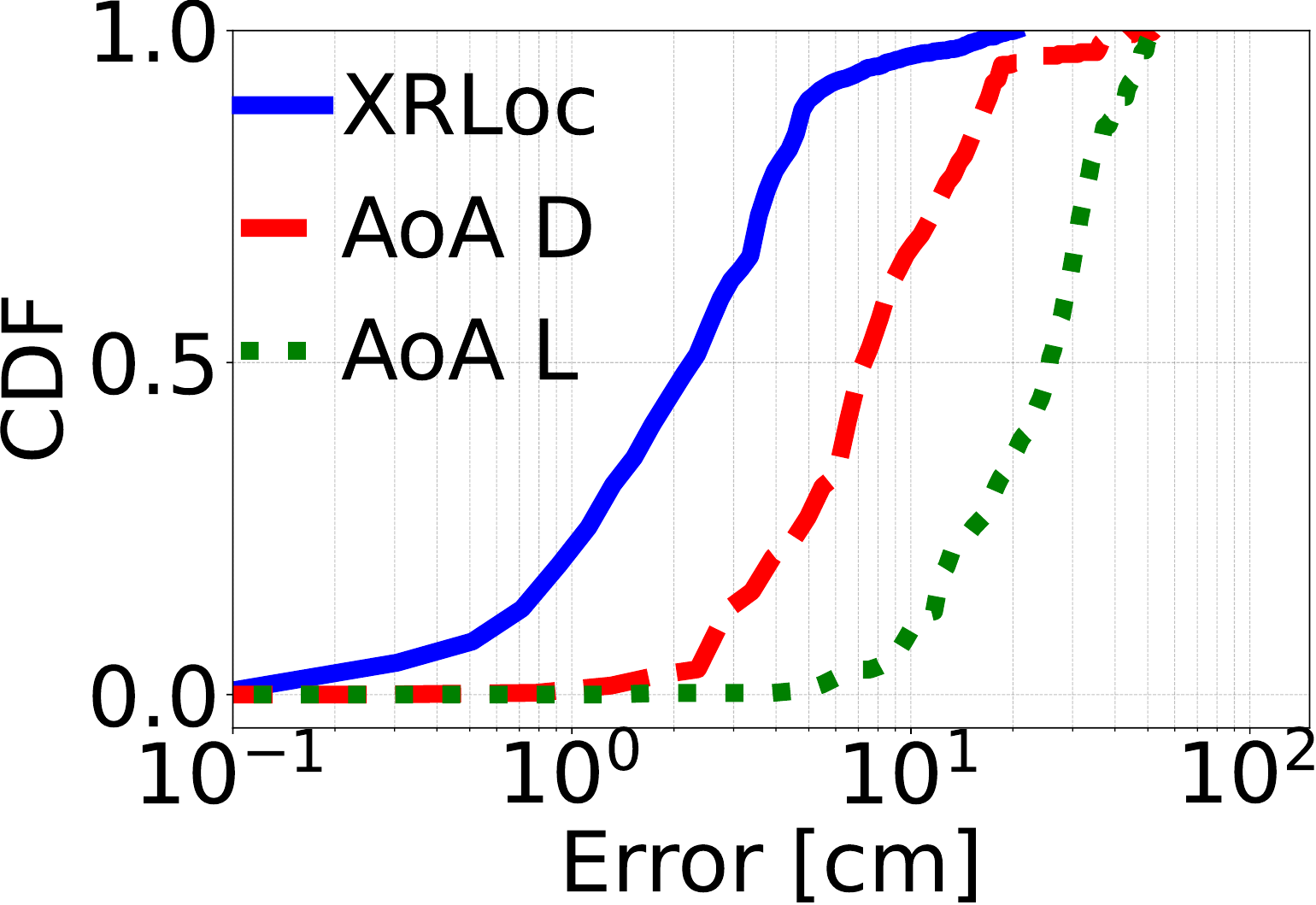}
            \caption{}
    \end{subfigure}
    ~
    \begin{subfigure}{0.3\linewidth}
            \includegraphics[scale=0.67]{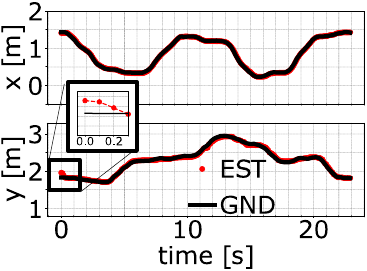}
            \caption{}
    \end{subfigure}
    \caption{Dynamic testing scenario: (a, b) Scatter plot of \name's predictions (EST), ground truth (GND), and antenna locations (ANT). More examples can be found in demo video$^{\ref{fn:demo}}$. (c) CDF of errors compared to AoA-based localization when three anchors~\cite{zhao2021uloc} are placed diversely around the room (AoA-D) and constrained to a $1$ m single line (AoA-L). (d) time-series errors of movement in (b), with the inset showcasing particle filter convergence within five packets.} 
    \label{fig:eval-moving}
\end{figure*}

\subsection{Static Localization Accuracy}\label{sec:static}

One of the key use cases targeted in \name is to provide accurate locations of real-world objects and place them in the virtual realm. These objects of interest could be tagged with inexpensive and long-lasting UWB tags, which will relay their location to the VR system. 
To simulate this use case, we place multiple tags in the environment with the simple goal of recreating a life-size chess game. 
In this static scenario, from Fig.~\ref{fig:eval-static}(a), we observe a median and $90^\mathrm{th}$ percentile error of $1.5$ cm and $5.5$ cm, respectively. 
We additionally observe \name provides a $9.5 \times$ and $4.0 \times$ improvement at the median over using ULoc in a linear (AoA-L) and diverse (AoA-D) placement scenario which have (median, $90^\mathrm{th}\%$) of (14.6 cm, 28.7 cm) and (6.1 cm, 13.7 cm), respectively. 
The evaluation of different ranges shows median errors of 6.8 cm and 15.2 cm at 4m and 5 m in the LOS condition, respectively, and 35.3 cm and 34.0 cm in the NLOS condition as shown in Fig.~\ref{fig:eval-static}(b).

\subsection{Moving Localization Accuracy}\label{sec:moving}

Continuing with the motivation of playing a life-size chess game, we characterize \name's localization accuracy in dynamic scenarios. Fig.~\ref{fig:eval-moving}(a) and ~\ref{fig:eval-moving}(b) showcase two characteristic movement patterns we tested. We tested $8$ movements, as shown in the demo video$^{\ref{fn:demo}}$, and achieved median and $90^\mathrm{th}$ errors of 2.4 cm and 5.3 cm, respectively, as shown in Fig.~\ref{fig:eval-moving}(c). We observe an $11 \times$ and $3.2 \times$ improvement at median over using ULoc in a linear (AoA-L) and diverse (AoA-D) placement scenario, which have (median, $90^\mathrm{th}\%$) of (26.0, 43.3  cm) and (7.5 cm, 17.4 cm), respectively. 

 In Fig.~\ref{fig:eval-moving}(d), we show the time-series error of localization for the `Fig.~\ref{fig:eval-moving}(b)' movement scenario (Fig.~\ref{fig:eval-moving}(c)). We note that opting to use a particle filter over a brute force approach provides a localization latency of $~1$ ms, compared to exhaustive grid search's latency of $~61.2$ s on a 12 Core CPU as explained in Sec.~\ref{sec:des-opt}. However, because the particle filter performs a sparse sampling over the entire space, \name may initialize the tag's location incorrectly. This is visible in the inset shown in Fig.~\ref{fig:eval-moving}(d). But, throughout $5$ received packets, we can see the location converges to the true location, and \name subsequently provides accurate location predictions.

\begin{figure}
    \begin{subfigure}{0.49\linewidth}
        \includegraphics[scale=0.16]{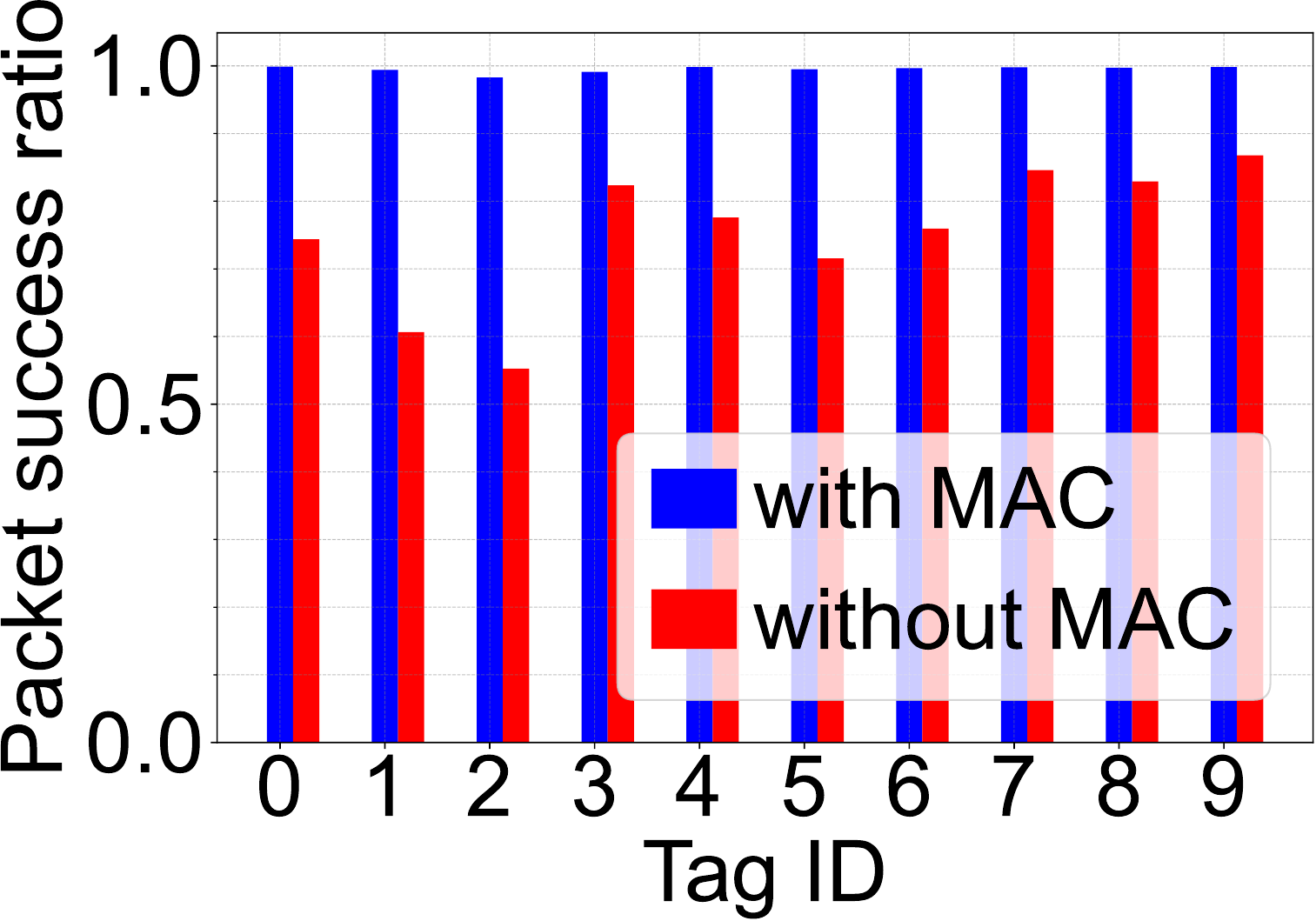}
        \caption{}
    \end{subfigure}
    ~
    \begin{subfigure}{0.49\linewidth}
        \includegraphics[scale=0.16]{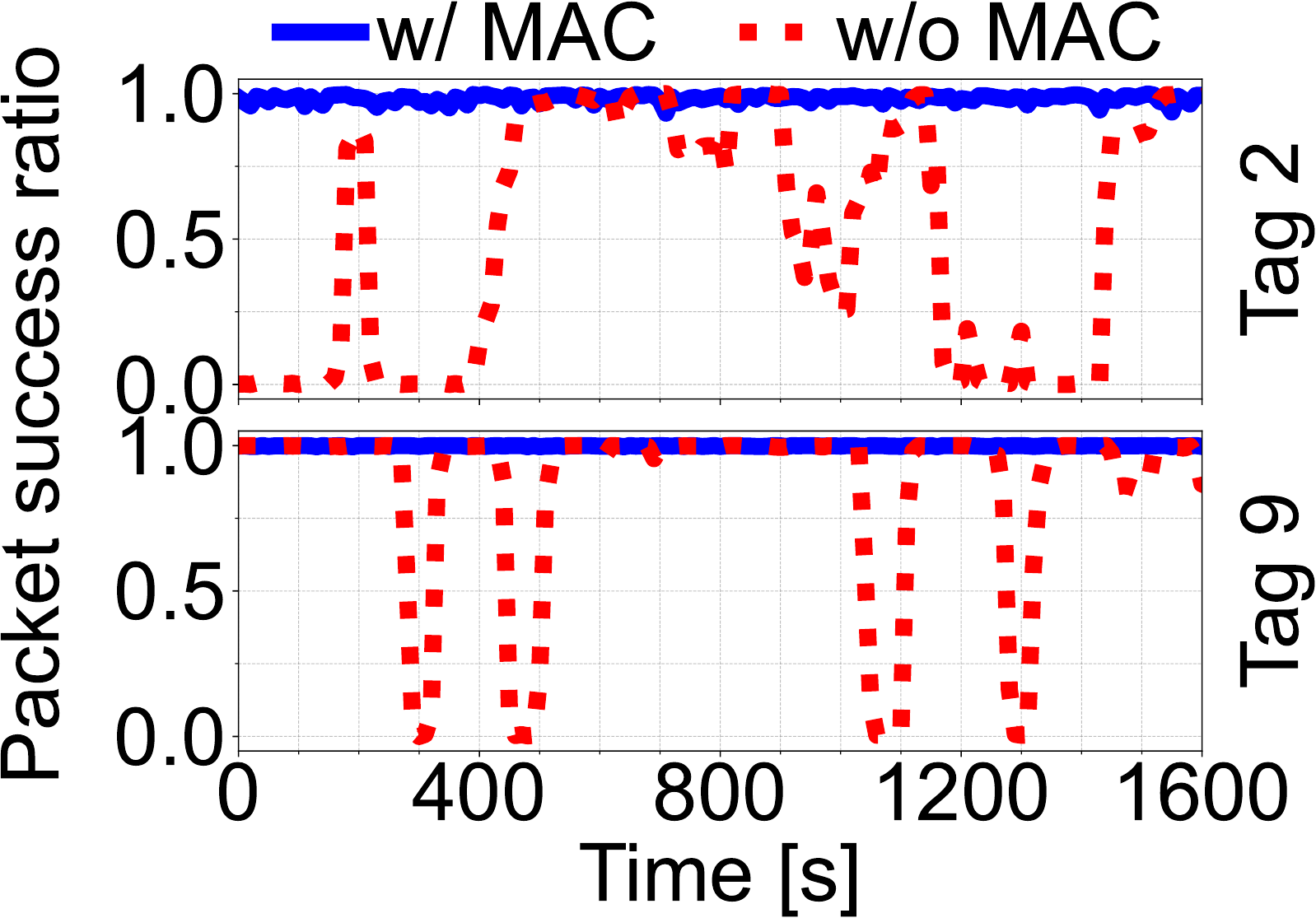}
        \caption{}
    \end{subfigure}
    \caption{MAC protocol performance: (a) Packet success ratio across ten tags with (blue) and without (red) mac protocol. (b) packet success ratio over time for Tag 2 (best performing \textit{without} MAC) and Tag 9 (worst performing \textit{without} MAC). In all cases, MAC protocol provides a success rate of over $99.5\%$.}
    \label{fig:eval-mac}
\end{figure}

\subsection{MAC Protocol Efficacy}

In the previous sections, we have shown \name can achieve a few-cm level localization from a single localization module, meeting the first two requirements (\textbf{R1} and \textbf{R2}). To allow multiple tags to be localized with this accuracy, \name leverages a LoRa side-channel to develop a power-efficient MAC protocol as described in Sec.~\ref{sec:des-mac}. To evaluate its efficacy, we set up $10$ tags to transmit at $100$ Hz for a half-hour period. Fig.~\ref{fig:eval-mac}(a) showcases the packet success ratio, and we find over $99.5\%$ of the packets are received by \name's localization module. Alternatively, when we do not have a MAC protocol, we have an average success rate of $~76\%$, ranging between $56\% - 87\%$. Specifically, considering the best and worst tag, we plot the packet arrival rate in Fig.~\ref{fig:eval-mac}(b) over the $30$ min period and observe there are large periods when packets from Tag 09 are not received, likely due to collision from either Tag 02 or any of the other tags in the environment. Alternatively, we see a consistent packet arrival rate using a MAC protocol. Clearly, a MAC protocol is necessary to achieve multi-tag tracking and localization at high rates and fulfill \textbf{R3}.

\begin{figure*}    
    \centering
    \newcommand{\mbscale}{0.16}
    \begin{subfigure}{0.25\linewidth}
        \includegraphics[scale=\mbscale]{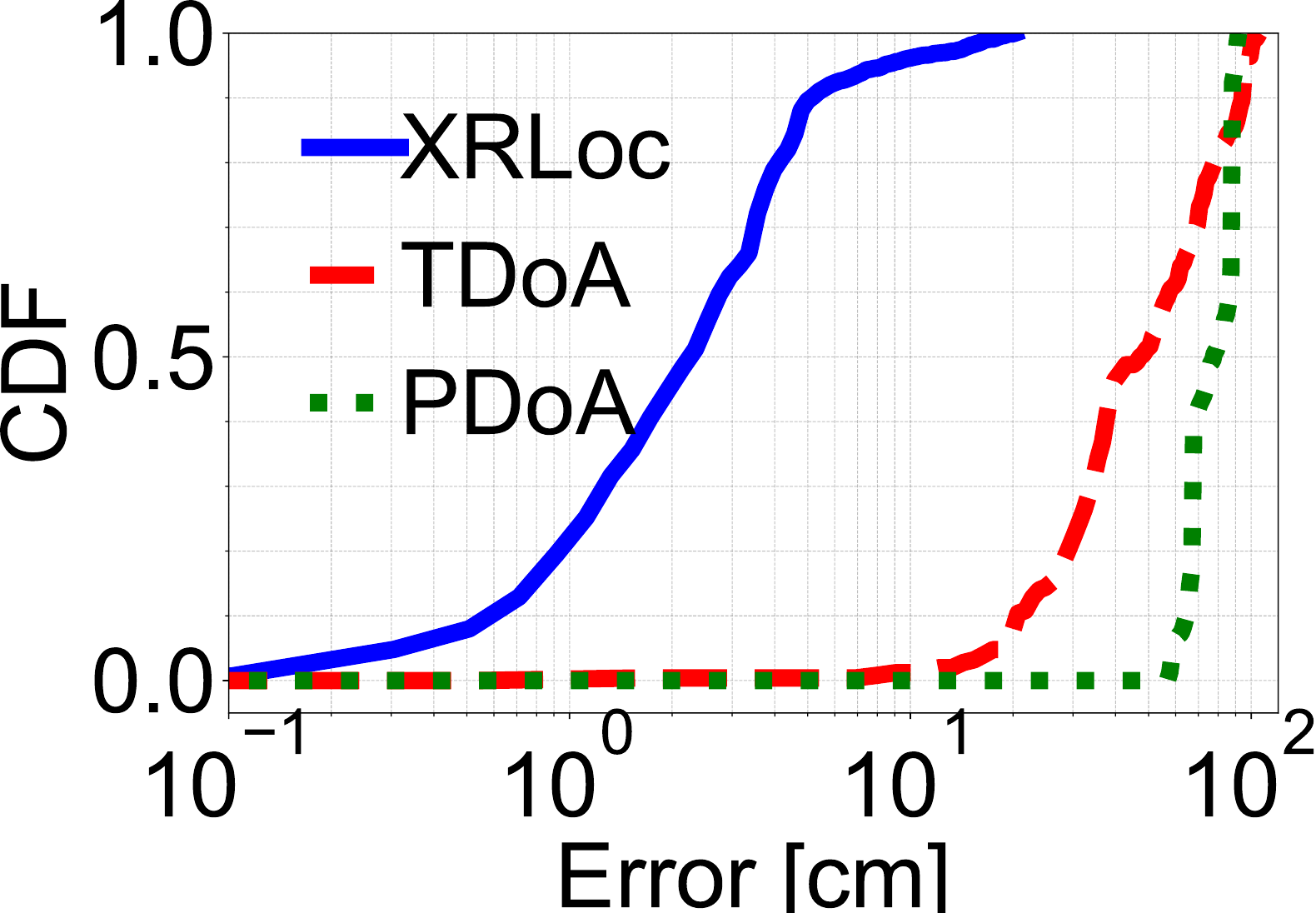}
        \caption{TDoA/PDoA}
    \end{subfigure}
    ~
    \begin{subfigure}{0.24\linewidth}
        \includegraphics[scale=\mbscale]{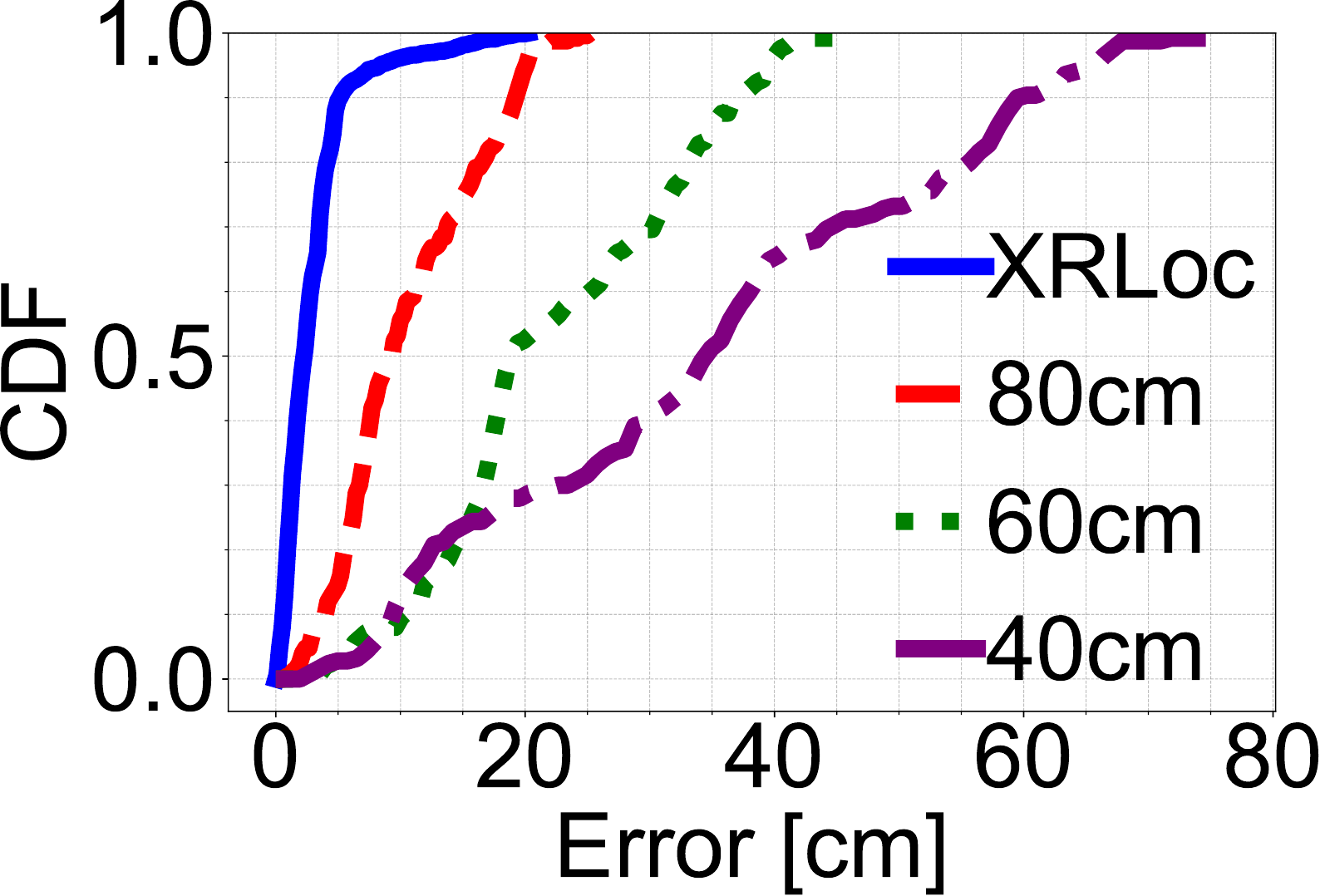}
        \caption{Aperture}
    \end{subfigure}
    ~
    \begin{subfigure}{0.25\linewidth}
        \includegraphics[scale=\mbscale]{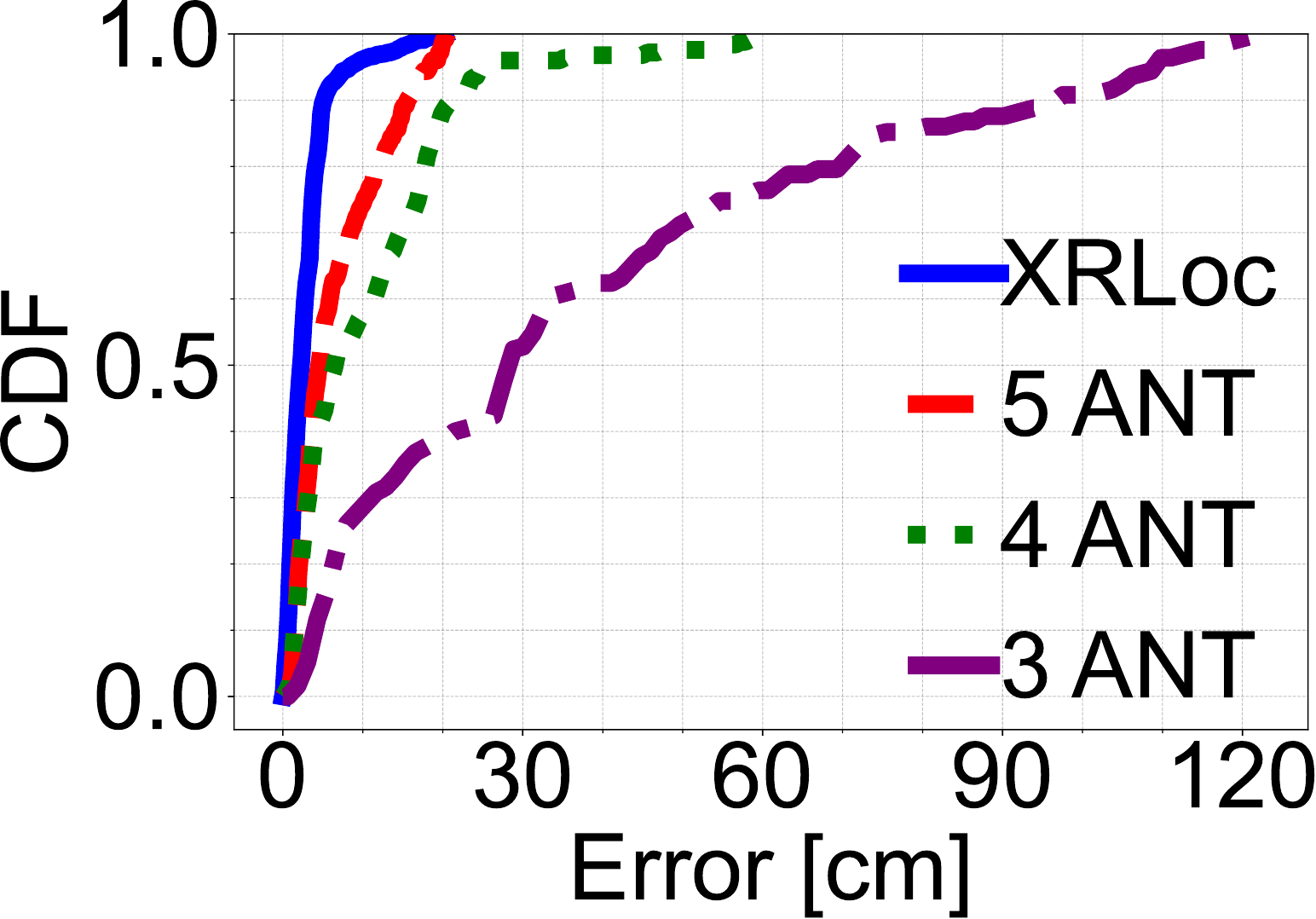}
        \caption{\# of antennas}
    \end{subfigure}
    ~
    \begin{subfigure}{0.25\linewidth}
        \includegraphics[scale=\mbscale]{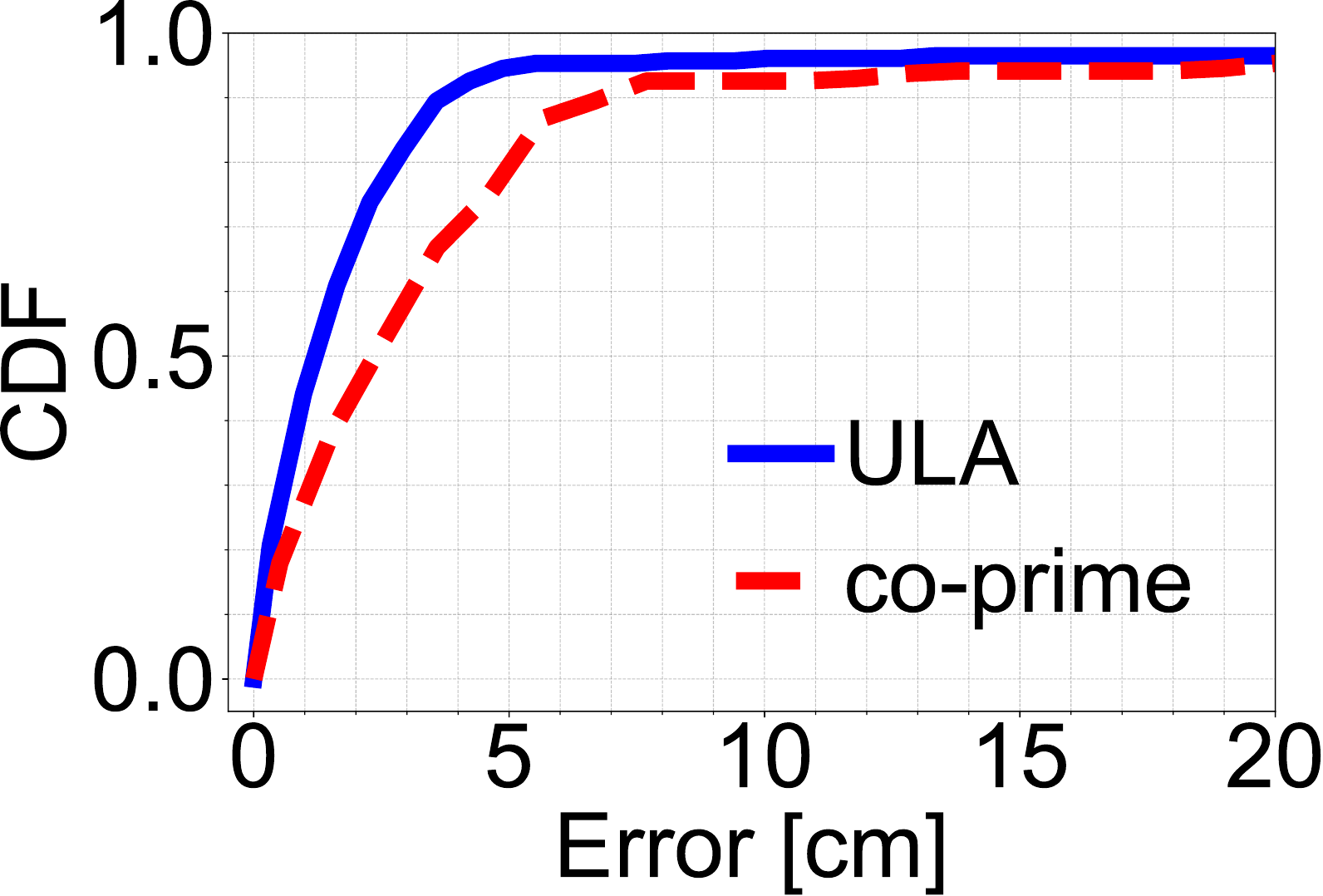}
        \caption{Antenna patterns}
    \end{subfigure}
    \caption{Microbenchmarks: (a) Using TDoA or PDoA only as opposed to a fusion (\name). (b) Reducing aperture from $1$ m (\name). (c) Reducing the number of antennas while keeping the aperture at $1$ m. (d) Leveraging co-prime antenna array as opposed to uniform linear array (ULA).}
    \label{fig:eval-mb}
\end{figure*}

\subsection{Justifying design choices}\label{sec:mb}

The evaluations from the previous sections prove \name's ability to fulfill the stringent requirements set for Sec.~\ref{sec:introduction}. In the following section, we will answer key questions about the design choices made when developing \name.

\noindent \textbf{TDoA and PDoA are both needed?:}
As we have discussed, a system relying purely on time-based measurements will not meet the stringent requirements of few-cm localization accuracy. We further evaluate this on our datasets in Fig.~\ref{fig:eval-mb}(a). We see a median localization accuracy of 2.4 cm, deviating over an order of magnitude from our few-cm level accuracy requirement. This re-iterates the challenge of achieving single-vantage point localization. 

However, we claimed in Sec.~\ref{sec:des-amb} TDoA measurements play an important role in ruling out ambiguous initialization caused by PDoA-only localization. To confirm this, we see in the same figure when PDoA is solely used for localization, and we have a median accuracy of 49.1 cm. Clearly, ambiguities from phase wrap-around can be detrimental to \name's performance, emphasizing TDoA's role. Through this micro-benchmark, it is apparent TDoA and PDoA work hand-in-hand to provide few-cm location accuracy.   

\noindent \textbf{How does the aperture effect the localization?:}
In Sec.~\ref{sec:des-res}, we discussed the importance of the antenna aperture in bringing resilience to phase measurement error. Consequently, a wider distance between the first and last antenna helps to improve localization accuracy. To ensure easy integration within everyday consumer electronics (like TVs or soundbars), we restrict \name's size to less than $1$ m wide. However, how important is antenna aperture to our localization performance? For this, we reduce the maximum antenna aperture to $80$, $60$, and $40$ cm and report the results in Fig.~\ref{fig:eval-mb}(b). Clearly, a reduction in the aperture size affects the localization accuracy, with median localization accuracy reducing to 9.6, 19.3, and 35.0 cm, respectively. In fact, we see a steep drop-off in accuracy when we have an aperture of $40$ cm. Furthermore, we see that a minimum aperture of $1$ m is required to achieve the required localization accuracy. Under space constraints, smaller apertures may be used at the cost of lower accuracy. 

\noindent \textbf{How many antennas are needed?:}
Clearly, a minimum aperture of $1$ m is needed. However, within this aperture, how many antennas are needed to meet the localization requirements? This is an important question to consider to make \name cost-effective. In the previous localization accuracy analysis, we consider an array with 6 antennas. In Fig.~\ref{fig:eval-mb}(c), we reduce the number of antennas placed within the $1$ m aperture. For 6, 5, and 4 antennas, we see the median location accuracy of 4.7, 6.9, and 28.7 cm, respectively. As few as 4 antennas are enough to meet the required few-cm localization accuracy at the median. Although, we observe a sharp reduction in localization accuracy in the $90^\mathrm{th}$ percentile. More antennas provide a better averaging effect and reduce erroneous TDoA and PDoA measurements, hence improving the localization performance at higher percentiles. From these experiments, we empirically observe choosing at least $6$ antennas meets the required few-cm level accuracy required for XR applications.  

\noindent \textbf{Are there better antenna spacing we can choose?:}
So far, we have considered placing our antennas in a uniform linear array (ULA), separated by $20$ cm. However, many works~\cite{vaidyanathan2010sparse, wang2014rf} showcase antenna patterns that are more optimal than a ULA. To investigate the improvements from these co-prime antenna arrays, we leverage our simulator from Sec.~\ref{sec:back} to carry out extensive simulations and showcase the results in Fig.~\ref{fig:eval-mb}(d). We see slight degradation of error when using co-prime arrays. However, co-prime arrays can be levered to reduce the number of antennas required by \name to achieve similar location accuracy.

\begin{figure}
    \begin{subfigure}{0.5\linewidth}
        \includegraphics[scale=0.15]{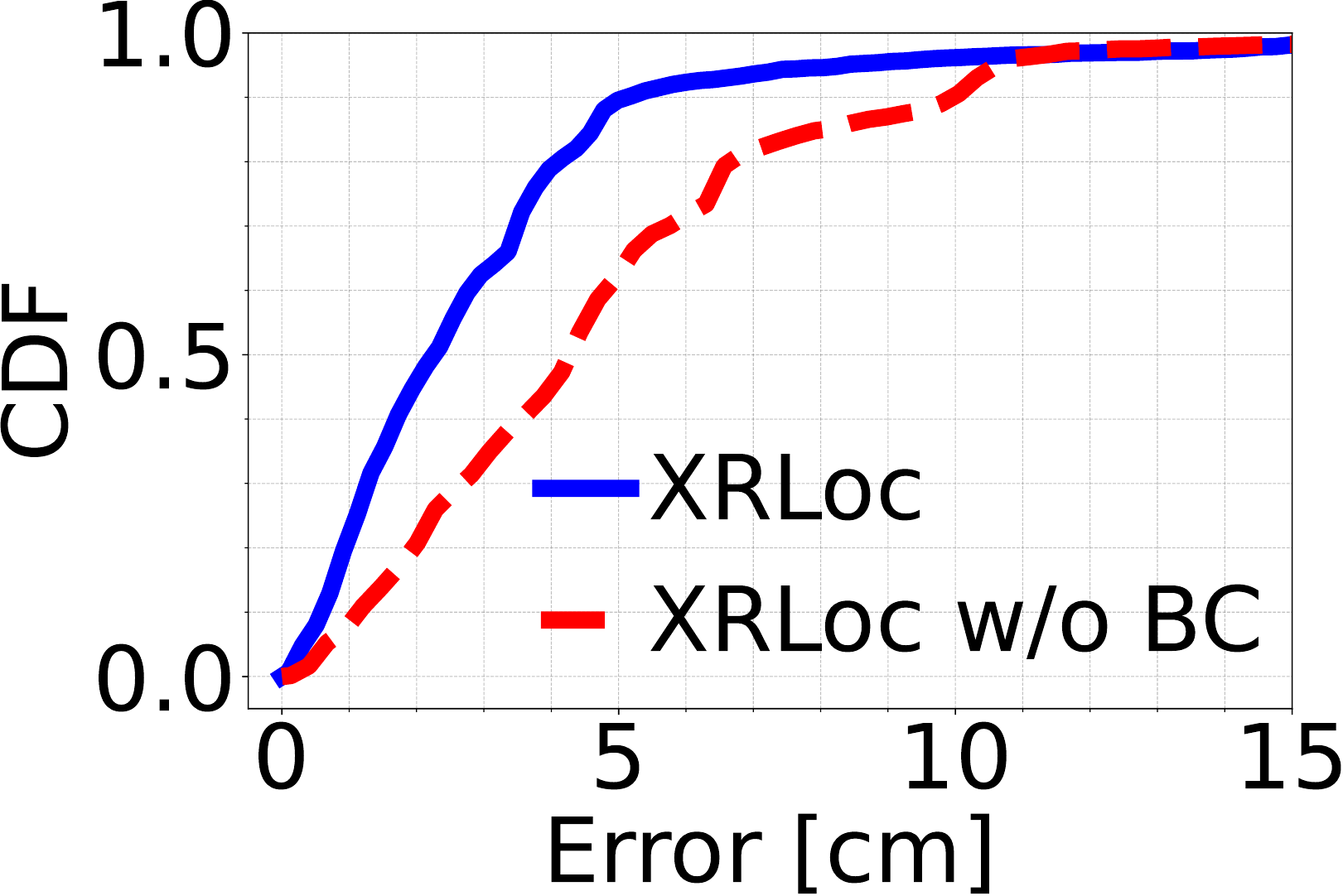}
        \caption{}
    \end{subfigure}
    ~
    \begin{subfigure}{0.5\linewidth}
        \includegraphics[scale=0.14]{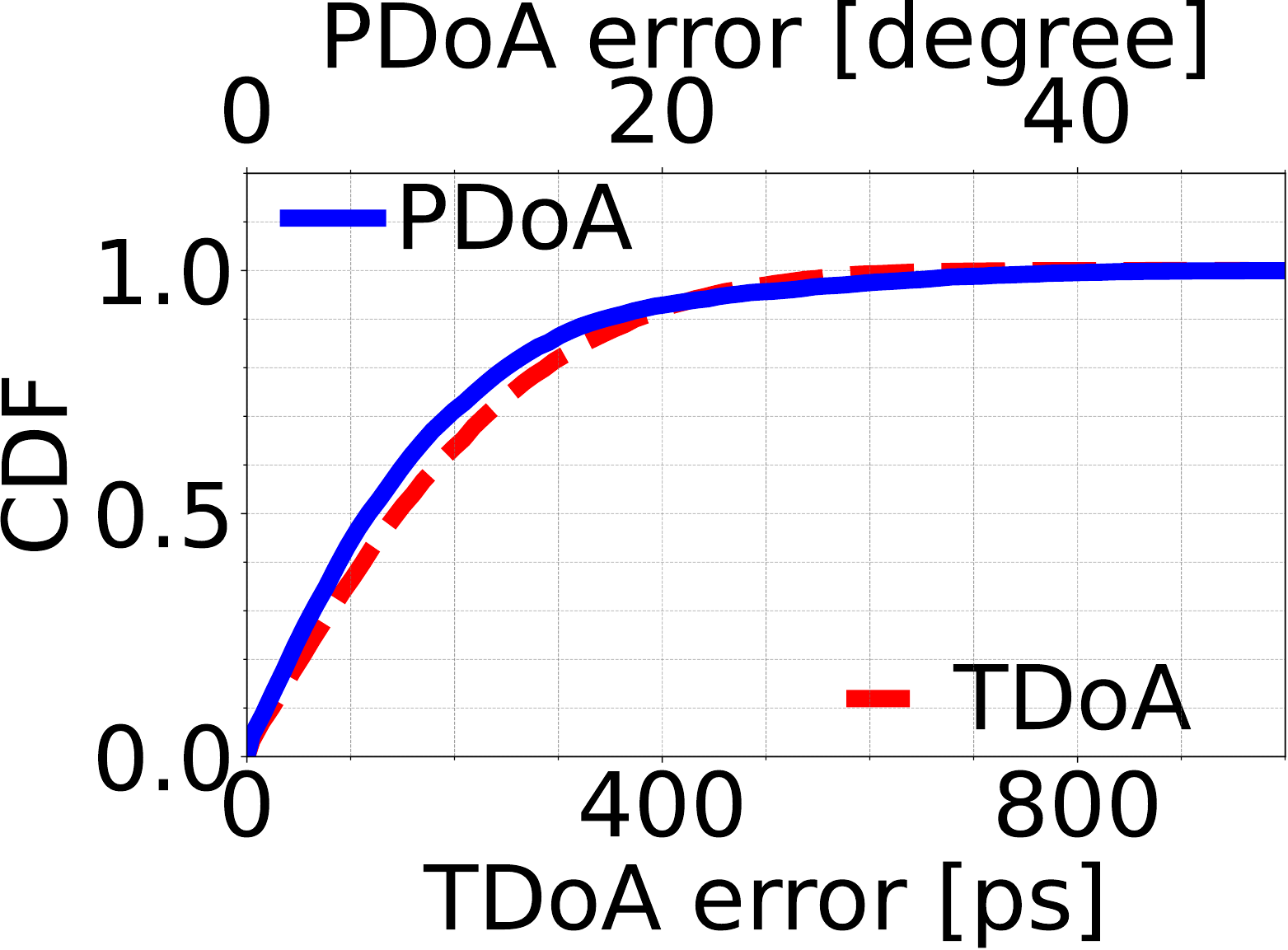}
        \caption{}
    \end{subfigure}
    \caption{(a) Localization error with and without bias calibration; (b) measured PDoA and TDoA errors.}
    \label{fig:eval-hw}
\end{figure}

\noindent \textbf{Why do we need fine-grained bias compensation?:}
Finally, we evaluate the system-level measurements. In \name, we choose the appropriate clock sources to achieve the required accuracy in both TDoA and PDoA measurements (Sec.~\ref{im:clock}) and additionally calibrate for TDoA and PDoA hardware biases via a 3-point calibration scheme (Sec.~\ref{im:calib}). In Fig.~\ref{fig:eval-hw}(a), we showcase the importance of this bias calibration, observing median localization accuracy degrade by $1.8 \times$ to a median accuracy of 2.4 cm without applying appropriate bias calibration. In Fig.~\ref{fig:eval-hw}(b), we also observe an average TDoA error of $180.7$ ps and PDoA error of $8.2^\circ$.

\section{Related Works}
\label{sec:related}
Providing indoor location information for people and various in-animate objects is a well-studied problem. This section will broadly cover various techniques leveraged to address this problem. We will find that none of the existing techniques meets the stringent requirements we set up earlier in Sec.~\ref{sec:introduction}. Recall that we seek to provide easy-to-deploy (\textbf{R1}), few-cm accurate localization (\textbf{R2}) in dynamic scenarios for multiple people or objects of interest (\textbf{R3}). A few key technologies which can be considered are:

\noindent \textbf{Visual sensing:} Under this broad umbrella, we have many distinct technologies. Existing VR systems utilize external IR-based sensors~\cite{borges2018htc} or specialized cameras~\cite{vicon} to furnish accurate ground truth locations. There are also works that deploy a single Lidar~\cite{hasan2022lidar} for person tracking or utilize headset-mounted cameras~\cite{monica2022evaluation}. However, these systems are sensitive to visual occlusions, hindering a user experience. Recent works~\cite{zhong2021towards, sahin2023hoot, luo2022novel} which leverage machine learning to track objects despite occlusions. 
Alternatively, other studies ~\cite{guo2022multi, li2020pose, sensys2020litag,sensys2021curvelight,imwut2023spectral} seek to deploy multiple cameras, let tag equips with a camera, or utilize special light sources to be robust to occlusions.
However, no studies have simultaneously solved all the problems of ease of anchor deployment (\textbf{R1}), accuracy (\textbf{R2}), and the risk of security and privacy~\cite{vigdor2019somebody}.
Moving away from deploying privacy-invasive cameras, other works~\cite{miller2022cappella} seek to use the cameras on-board VR setups fused with occlusion-resilient radio-frequency (RF) signals like ultra-wideband.
These systems have a low deployment cost but do not achieve a few-cm level accuracy. 

\noindent \textbf{Acoustic sensing:}  Alternative to these systems, various acoustic localization systems~\cite{famili2022pilot, merenda2022rfid, murakami20193,cao2020earphonetrak,sensys2020fmtrack,sensys2020symphony} take advantage of the lower speed of sound (~$340$ m/s) for fine-grained localization and meet the required localization accuracy. However, acoustic sensing has a few fundamental drawbacks~\cite{li2022experience}. 
First, acoustic systems~\cite{liu2020indoor} are difficult to provide both multi-asset and low latency localization simultaneously because of narrow bandwidth, deviating from \textbf{R3}.
Second, they hinder music and audio playback, precluding immersive  XR applications. Third, acoustic signals that employ ultrasound (> $20$ kHz) for sensing have considerable audio leakage in the audible frequency range, affecting user experience. 

\noindent \textbf{Radar-based sensing:} Mm-wave radars near the $60$ GHz and $77$ GHz bands have gained recent interest. Many works~\cite{xue2021mmmesh, kong2022m3track} have looked at furnishing human pose with these radars from a single radar. Recent work~\cite{mukherjee2022scalable} has shown that the human body can act as a strong blockage at these frequencies. These blockages can hinder tracking multiple people and objects in an environment and affect user experience. Additionally, tracking and identifying smaller assets in an environment can be challenging as radar reflections depend on an object's radar cross-sectional area (RCS). Alternatively, many works~\cite{soltanaghaei2021millimetro} propose placing retro-reflective tags on objects with small RCS to guarantee their detection; however, these systems suffer from poor localization accuracy. 

\noindent \textbf{RF-based sensing:} 
The robustness of sub-6 GHz RF-signals to occlusions~\cite{slezak2022measurement} and low privacy risk makes it a promising technology to consider. The common mode of operation is for multiple RF radios to jointly localize an active RF transmitter or a passive RF reflector (tags). Many works have looked at leveraging WiFi~\cite{mostafa2022survey, spotfi, arraytrack, vasisht2016decimeter}, LoRa~\cite{percom2019lorain}, or BLE~\cite{ayyalasomayajula2018bloc} to achieve robust user localization. However, these systems fail to provide the required localization accuracy due to bandwidth limitations. 

RFID has a strong asymmetry in the reader-tag relationship, and the transmitter and receiver share the same clock, which allows for highly accurate phase acquisition.
According to \cite{mobitagbot,tagoram,turbotrack}, RFID systems do not have carrier and sampling frequency offset and enjoy a phase measurement accuracy of $0.085^\circ$~\cite{tagoram}, 15$\times$ better than the UWB, which provides an accuracy of ~$1.4^\circ$.
Using the highly accurate phase, \cite{tagoram,turbotrack,wang2014rf,ma2017minding,ipsn2018omnitrack} has succeeded with tracking or localization at the few cm levels.
However, due to the asymmetric nature, RFID readers whose range is several meters are not suitable for embedding into consumer electronics (\textbf{R1}) because of their power-hungriness and expensiveness (ex. ImpinjJ Speedway R420 costs \$1666).
The main target of RFID is industrial or retail store settings where thousands of tags must be deployed inexpensively, and readers' one-time cost is justifiable.
For instance, \cite{mobitagbot} looks at item ordering in manufacturing lines, retail stores, or libraries. \cite{tagoram,turbotrack} examine industrial robotics or baggage handling tasks.

Unlike RFID, Ultra-wideband provides a more symmetric architecture where localization modules can cost $\$10-100$. Consequently, we have seen their increased adoption in smartphones and smart tags. It provides over $500$ MHz of bandwidth and a time resolution of $1$ ns, providing localization accuracy to a few tens of centimeters.
Many current UWB-localization schemes leverage the accurate time-resolution for Two-Way Ranging (TWR)~\cite{zwirello2012uwb,ledergerber2019ultra,garcia2015robust,alarifi2016ultra,bonnin2019uwb, kempke2016harmonium,9289338,iball} and localize objects via trilateration. However, these multiple-packet exchanges increase localization latency and prevent real-time tracking of multiple objects of interest (\textbf{R3}). Many works instead leverage the TDoA or PDoA of the UWB signal to multiple time-synchronized anchors~\cite{chen2020joint, vecchia2019talla, tiemann2016atlas, grobetawindhager2019snaploc,cao2021itracku}, or AoA measurements~\cite{zhao2021uloc, heydariaan2020anguloc, decawave-pdoa-kit} at multiple anchors to furnish locations using a single packet. Some works~\cite{yang2022vuloc} employ alternative transmission schemes to TWR to reduce the packet overhead. However, these systems only meet the necessary localization accuracy when the UWB anchors are placed in diverse locations, increasing deployment efforts and deviating from \textbf{R1}.

As discussed in Sec.~\ref{sec:back}, few-cm accurate localization is challenging due to geometric dilution of precision. To circumvent this problem, three common techniques are leveraged. First, by leveraging reflected paths in the environment, many systems~\cite{kotaru2017position, chen2019m, li2020multipath, zhang2022toward, soltanaghaei2018multipath, meissner2012multipath, grosswindhager2018salma} create additional ``virtual'' radios in the environment. These ``virtual'' radios provide the needed spatial diversity to localize an object of interest. However, multipath is often unreliable~\cite{almers2006keyhole} in many environments and can lead to localization failure and poor user experience. Second, many works~\cite{wang2019high, ge2021single, wang2019efficient, rostami2022enabling} look at fusing TWR, TDOA, and AoA information to provide single anchor localization solutions. However, some systems cannot furnish the few-cm accurate localization requirement or rely on TWR measurements, increasing the system's latency. Finally, some works develop switched beam antennas~\cite{groth2021calibration, giorgetti2009single}, which selectively sense signals approaching the anchor from different directions. However, these systems lack the required angular resolution to provide localization accuracy of a few cm.

\section{Discussion and Future work}

\name overcomes the fundamental challenges arising from geometric dilution of precision to deliver cm-level accurate localization by developing an easy-to-deploy and low-latency localization module. Through this development, we are one step closer to achieving immersive XR experiences. However, a few limitations and possibilities of future work can be explored to build upon \name.

\noindent \textbf{Extensions to 3D:} \name focuses on localizing people and assets on a 2D floor plane, which is required in various XR applications. However, these ideas can be extended to the 3D domain by incorporating a vertical array of antennas in conjunction with the current horizontal linear array. These 3D-compliant antenna arrays can be retrofitted with television screens or paintings to allow cm-accurate 3D localization. 

\noindent \textbf{Improving power efficiency of \name's localization module:} Various works~\cite{biri2020socitrack} have noted the $10 \times$ higher power consumption of UWB reception than transmission. Keeping this in mind, we designed a system that requires only a single transmission from the tag for localization to ensure long battery life. However, the $6$ receivers on \name's wall-powered localization module are power inefficient. To rectify this, antenna switching schemes~\cite{gu2021tyrloc}  can be employed, or multiple antennas can be combined to connect to a single receiver~\cite {cao2021itracku} to reduce the number of receivers. However, unlike \name's system, these alternatives will not be FiRa compliant~\cite{coppens2022overview}.

\noindent \textbf{Miniaturized tag design: } We prototype our tag from off-the-shelf EVB1000~\cite{evb1000} and LoRa~\cite{sx1272mb2das} evaluation boards. Future work can look towards miniaturizing these tag designs. Since these radios we employ are centered at 3.4 GHz and 930 MHz, it allows us to place these radio modules in close proximity with limited RF interference. 

\section{Acknowledgement}

We thank Neil Smith at UCSD, Kazuhiro Kizaki at Osaka University, and the members of WCSNG at UCSD for their help and feedback.

\newpage
\bibliographystyle{ACM-Reference-Format}
\bibliography{references2}


\begin{thebibliography}{103}


\ifx \showCODEN    \undefined \def \showCODEN     #1{\unskip}     \fi
\ifx \showDOI      \undefined \def \showDOI       #1{#1}\fi
\ifx \showISBNx    \undefined \def \showISBNx     #1{\unskip}     \fi
\ifx \showISBNxiii \undefined \def \showISBNxiii  #1{\unskip}     \fi
\ifx \showISSN     \undefined \def \showISSN      #1{\unskip}     \fi
\ifx \showLCCN     \undefined \def \showLCCN      #1{\unskip}     \fi
\ifx \shownote     \undefined \def \shownote      #1{#1}          \fi
\ifx \showarticletitle \undefined \def \showarticletitle #1{#1}   \fi
\ifx \showURL      \undefined \def \showURL       {\relax}        \fi
\providecommand\bibfield[2]{#2}
\providecommand\bibinfo[2]{#2}
\providecommand\natexlab[1]{#1}
\providecommand\showeprint[2][]{arXiv:#2}

\bibitem[{Abracon LLC}(2022)]%
        {astxr}
\bibfield{author}{\bibinfo{person}{{Abracon LLC}}.}
  \bibinfo{year}{2022}\natexlab{}.
\newblock \bibinfo{title}{ASTXR-12-38.400MHZ-514054-T}.
\newblock
  \bibinfo{howpublished}{\url{https://abracon.com/Oscillators/ASTXR-12-38.400MHz-514054-T.pdf}}.
\newblock
\newblock
\shownote{Accessed: 2023-02-14}.


\bibitem[Aernouts et~al\mbox{.}(2020)]%
        {aernouts2020combining}
\bibfield{author}{\bibinfo{person}{Michiel Aernouts}, \bibinfo{person}{Noori
  BniLam}, \bibinfo{person}{Nico Podevijn}, \bibinfo{person}{David Plets},
  \bibinfo{person}{Wout Joseph}, \bibinfo{person}{Rafael Berkvens}, {and}
  \bibinfo{person}{Maarten Weyn}.} \bibinfo{year}{2020}\natexlab{}.
\newblock \showarticletitle{Combining TDoA and AoA with a particle filter in an
  outdoor LoRaWAN network}. In \bibinfo{booktitle}{\emph{IEEE/ION Position,
  Location and Navigation Symposium (PLANS)}}. \bibinfo{pages}{1060--1069}.
\newblock


\bibitem[Alarifi et~al\mbox{.}(2016)]%
        {alarifi2016ultra}
\bibfield{author}{\bibinfo{person}{Abdulrahman Alarifi},
  \bibinfo{person}{AbdulMalik Al-Salman}, \bibinfo{person}{Mansour Alsaleh},
  \bibinfo{person}{Ahmad Alnafessah}, \bibinfo{person}{Suheer Al-Hadhrami},
  \bibinfo{person}{Mai Al-Ammar}, {and} \bibinfo{person}{Hend Al-Khalifa}.}
  \bibinfo{year}{2016}\natexlab{}.
\newblock \showarticletitle{Ultra wideband indoor positioning technologies:
  Analysis and recent advances}.
\newblock \bibinfo{journal}{\emph{Sensors}} \bibinfo{volume}{16},
  \bibinfo{number}{5} (\bibinfo{year}{2016}), \bibinfo{pages}{707:1--36}.
\newblock


\bibitem[Alatise and Hancke(2020)]%
        {alatise2020review}
\bibfield{author}{\bibinfo{person}{Mary Alatise} {and} \bibinfo{person}{Gerhard
  Hancke}.} \bibinfo{year}{2020}\natexlab{}.
\newblock \showarticletitle{A review on challenges of autonomous mobile robot
  and sensor fusion methods}.
\newblock \bibinfo{journal}{\emph{IEEE Access}}  \bibinfo{volume}{8}
  (\bibinfo{year}{2020}), \bibinfo{pages}{39830--39846}.
\newblock


\bibitem[Alizadehsalehi et~al\mbox{.}(2020)]%
        {alizadehsalehi2020bim}
\bibfield{author}{\bibinfo{person}{Sepehr Alizadehsalehi},
  \bibinfo{person}{Ahmad Hadavi}, {and} \bibinfo{person}{Joseph~Chuenhuei
  Huang}.} \bibinfo{year}{2020}\natexlab{}.
\newblock \showarticletitle{From BIM to extended reality in AEC industry}.
\newblock \bibinfo{journal}{\emph{Automation in Construction}}
  \bibinfo{volume}{116} (\bibinfo{year}{2020}), \bibinfo{pages}{103254:1--13}.
\newblock


\bibitem[Almers et~al\mbox{.}(2006)]%
        {almers2006keyhole}
\bibfield{author}{\bibinfo{person}{Peter Almers}, \bibinfo{person}{Fredrik
  Tufvesson}, {and} \bibinfo{person}{Andreas Molisch}.}
  \bibinfo{year}{2006}\natexlab{}.
\newblock \showarticletitle{Keyhole effect in MIMO wireless channels:
  Measurements and theory}.
\newblock \bibinfo{journal}{\emph{IEEE Transactions on Wireless
  Communications}} \bibinfo{volume}{5}, \bibinfo{number}{12}
  (\bibinfo{year}{2006}), \bibinfo{pages}{3596--3604}.
\newblock


\bibitem[Ayyalasomayajula et~al\mbox{.}(2018)]%
        {ayyalasomayajula2018bloc}
\bibfield{author}{\bibinfo{person}{Roshan Ayyalasomayajula},
  \bibinfo{person}{Deepak Vasisht}, {and} \bibinfo{person}{Dinesh Bharadia}.}
  \bibinfo{year}{2018}\natexlab{}.
\newblock \showarticletitle{BLoc: CSI-based accurate localization for BLE
  tags}. In \bibinfo{booktitle}{\emph{ACM International Conference on emerging
  Networking EXperiments and Technologies (CoNEXT)}}.
  \bibinfo{pages}{126--138}.
\newblock


\bibitem[Bauwens et~al\mbox{.}(2021)]%
        {bauwens2021uwb}
\bibfield{author}{\bibinfo{person}{Jan Bauwens}, \bibinfo{person}{Nicola
  Macoir}, \bibinfo{person}{Spilios Giannoulis}, \bibinfo{person}{Ingrid
  Moerman}, {and} \bibinfo{person}{Eli De~Poorter}.}
  \bibinfo{year}{2021}\natexlab{}.
\newblock \showarticletitle{UWB-MAC: MAC protocol for UWB localization using
  ultra-low power anchor nodes}.
\newblock \bibinfo{journal}{\emph{Ad Hoc Networks}}  \bibinfo{volume}{123}
  (\bibinfo{year}{2021}), \bibinfo{pages}{102637:1--11}.
\newblock


\bibitem[Biri et~al\mbox{.}(2020)]%
        {biri2020socitrack}
\bibfield{author}{\bibinfo{person}{Andreas Biri}, \bibinfo{person}{Neal
  Jackson}, \bibinfo{person}{Lothar Thiele}, \bibinfo{person}{Pat Pannuto},
  {and} \bibinfo{person}{Prabal Dutta}.} \bibinfo{year}{2020}\natexlab{}.
\newblock \showarticletitle{SociTrack: Infrastructure-free interaction tracking
  through mobile sensor networks}. In \bibinfo{booktitle}{\emph{ACM Annual
  International Conference on Mobile Computing and Networking (MobiCom)}}.
  \bibinfo{pages}{1--14}.
\newblock


\bibitem[Bonnin-Pascual and Ortiz(2019)]%
        {bonnin2019uwb}
\bibfield{author}{\bibinfo{person}{Francisco Bonnin-Pascual} {and}
  \bibinfo{person}{Alberto Ortiz}.} \bibinfo{year}{2019}\natexlab{}.
\newblock \showarticletitle{An UWB-based System for Localization inside
  Merchant Vessels}. In \bibinfo{booktitle}{\emph{IEEE International Conference
  on Emerging Technologies and Factory Automation (ETFA)}}.
  \bibinfo{pages}{1559--1562}.
\newblock


\bibitem[Borges et~al\mbox{.}(2018)]%
        {borges2018htc}
\bibfield{author}{\bibinfo{person}{Miguel Borges}, \bibinfo{person}{Andrew
  Symington}, \bibinfo{person}{Brian Coltin}, \bibinfo{person}{Trey Smith},
  {and} \bibinfo{person}{Rodrigo Ventura}.} \bibinfo{year}{2018}\natexlab{}.
\newblock \showarticletitle{HTC Vive: Analysis and accuracy improvement}. In
  \bibinfo{booktitle}{\emph{IEEE/RSJ International Conference on Intelligent
  Robots and Systems (IROS)}}. \bibinfo{pages}{2610--2615}.
\newblock


\bibitem[Cao et~al\mbox{.}(2020)]%
        {cao2020earphonetrak}
\bibfield{author}{\bibinfo{person}{Gaoshuai Cao}, \bibinfo{person}{Kuang Yuan},
  \bibinfo{person}{Jie Xiong}, \bibinfo{person}{Panlong Yang},
  \bibinfo{person}{Yubo Yan}, \bibinfo{person}{Hao Zhou}, {and}
  \bibinfo{person}{Xiang-Yang Li}.} \bibinfo{year}{2020}\natexlab{}.
\newblock \showarticletitle{EarphoneTrack: Involving earphones into the
  ecosystem of acoustic motion tracking}. In \bibinfo{booktitle}{\emph{ACM
  Conference on Embedded Networked Sensor Systems (SenSys)}}.
  \bibinfo{pages}{95--108}.
\newblock


\bibitem[Cao et~al\mbox{.}(2021)]%
        {cao2021itracku}
\bibfield{author}{\bibinfo{person}{Yifeng Cao}, \bibinfo{person}{Ashutosh
  Dhekne}, {and} \bibinfo{person}{Mostafa Ammar}.}
  \bibinfo{year}{2021}\natexlab{}.
\newblock \showarticletitle{ITrackU: Tracking a pen-like instrument via UWB-IMU
  fusion}. In \bibinfo{booktitle}{\emph{ACM International Conference on Mobile
  Systems, Applications, and Services (MobiSys)}}. \bibinfo{pages}{453--466}.
\newblock


\bibitem[Chen et~al\mbox{.}(2020)]%
        {chen2020joint}
\bibfield{author}{\bibinfo{person}{Hui Chen}, \bibinfo{person}{Tarig Ballal},
  \bibinfo{person}{Nasir Saeed}, \bibinfo{person}{Mohamed-Slim Alouini}, {and}
  \bibinfo{person}{Tareq Al-Naffouri}.} \bibinfo{year}{2020}\natexlab{}.
\newblock \showarticletitle{A joint TDoA-PDoA localization approach using
  particle swarm optimization}.
\newblock \bibinfo{journal}{\emph{IEEE Wireless Communications Letters}}
  \bibinfo{volume}{9}, \bibinfo{number}{8} (\bibinfo{year}{2020}),
  \bibinfo{pages}{1240--1244}.
\newblock


\bibitem[Chen et~al\mbox{.}(2019)]%
        {chen2019m}
\bibfield{author}{\bibinfo{person}{Zhe Chen}, \bibinfo{person}{Guorong Zhu},
  \bibinfo{person}{Sulei Wang}, \bibinfo{person}{Yuedong Xu},
  \bibinfo{person}{Jie Xiong}, \bibinfo{person}{Jin Zhao}, \bibinfo{person}{Jun
  Luo}, {and} \bibinfo{person}{Xin Wang}.} \bibinfo{year}{2019}\natexlab{}.
\newblock \showarticletitle{M$^3$: Multipath assisted Wi-Fi localization with a
  single access point}.
\newblock \bibinfo{journal}{\emph{IEEE Transactions on Mobile Computing}}
  \bibinfo{volume}{20}, \bibinfo{number}{2} (\bibinfo{year}{2019}),
  \bibinfo{pages}{588--602}.
\newblock


\bibitem[Coppens et~al\mbox{.}(2022)]%
        {coppens2022overview}
\bibfield{author}{\bibinfo{person}{Dieter Coppens}, \bibinfo{person}{Eli
  De~Poorter}, \bibinfo{person}{Adnan Shahid}, \bibinfo{person}{Sam Lemey},
  {and} \bibinfo{person}{Chris Marshall}.} \bibinfo{year}{2022}\natexlab{}.
\newblock \showarticletitle{An overview of ultra-wideBand (UWB) standards (IEEE
  802.15. 4, FiRa, Apple): Interoperability aspects and future research
  directions}.
\newblock \bibinfo{journal}{\emph{{IEEE} Access}}  \bibinfo{volume}{10}
  (\bibinfo{year}{2022}), \bibinfo{pages}{70219--70241}.
\newblock


\bibitem[Corbal{\'a}n and Picco(2020)]%
        {corbalan2020ultra}
\bibfield{author}{\bibinfo{person}{Pablo Corbal{\'a}n} {and}
  \bibinfo{person}{Gian~Pietro Picco}.} \bibinfo{year}{2020}\natexlab{}.
\newblock \showarticletitle{Ultra-wideband concurrent ranging}.
\newblock \bibinfo{journal}{\emph{ACM Transactions on Sensor Networks}}
  \bibinfo{volume}{16}, \bibinfo{number}{4} (\bibinfo{year}{2020}),
  \bibinfo{pages}{1--41}.
\newblock


\bibitem[{Crystek Corp.}(2022)]%
        {crystek}
\bibfield{author}{\bibinfo{person}{{Crystek Corp.}}}
  \bibinfo{year}{2022}\natexlab{}.
\newblock \bibinfo{title}{Crystek XTAL OSC VCXO 122.8800MHZ}.
\newblock
  \bibinfo{howpublished}{\url{https://www.crystek.com/home/crystek/default.aspx}}.
\newblock
\newblock
\shownote{Accessed: 2023-02-14}.


\bibitem[De~Preter et~al\mbox{.}(2019)]%
        {de2019range}
\bibfield{author}{\bibinfo{person}{Andreas De~Preter}, \bibinfo{person}{Glenn
  Goysens}, \bibinfo{person}{Jan Anthonis}, \bibinfo{person}{Jan Swevers},
  {and} \bibinfo{person}{Goele Pipeleers}.} \bibinfo{year}{2019}\natexlab{}.
\newblock \showarticletitle{Range bias modeling and autocalibration of an UWB
  positioning system}. In \bibinfo{booktitle}{\emph{International Conference on
  Indoor Positioning and Indoor Navigation (IPIN)}}. \bibinfo{pages}{1--8}.
\newblock


\bibitem[Decawave(2020)]%
        {decawave-dw1000}
\bibfield{author}{\bibinfo{person}{Decawave}.} \bibinfo{year}{2020}\natexlab{}.
\newblock \bibinfo{title}{Datasheet: DW1000 IEEE802.15.4-2011 UWB
  Trans\-ceiver}.
\newblock
\newblock
\newblock
\shownote{Version 2.22}.


\bibitem[Deering(1998)]%
        {deering1998limits}
\bibfield{author}{\bibinfo{person}{Michael Deering}.}
  \bibinfo{year}{1998}\natexlab{}.
\newblock \showarticletitle{The limits of human vision}. In
  \bibinfo{booktitle}{\emph{International Immersive Projection Technology
  Workshop}}. \bibinfo{pages}{1--6}.
\newblock


\bibitem[Dotlic et~al\mbox{.}(2017)]%
        {decawave-pdoa-kit}
\bibfield{author}{\bibinfo{person}{Igor Dotlic}, \bibinfo{person}{Andrew
  Connell}, \bibinfo{person}{Hang Ma}, \bibinfo{person}{Jeff Clancy}, {and}
  \bibinfo{person}{Michael McLaughlin}.} \bibinfo{year}{2017}\natexlab{}.
\newblock \showarticletitle{Angle of arrival estimation using Decawave DW1000
  integrated circuits}. In \bibinfo{booktitle}{\emph{IEEE Workshop on
  Positioning, Navigation and Communications (WPNC)}}. \bibinfo{pages}{1--6}.
\newblock


\bibitem[Famili et~al\mbox{.}(2022)]%
        {famili2022pilot}
\bibfield{author}{\bibinfo{person}{Alireza Famili}, \bibinfo{person}{Angelos
  Stavrou}, \bibinfo{person}{Haining Wang}, {and}
  \bibinfo{person}{Jung-Min~Jerry Park}.} \bibinfo{year}{2022}\natexlab{}.
\newblock \showarticletitle{Pilot: High-precision indoor localization for
  autonomous drones}.
\newblock \bibinfo{journal}{\emph{IEEE Transactions on Vehicular Technology}}
  \bibinfo{volume}{72}, \bibinfo{number}{5} (\bibinfo{year}{2022}),
  \bibinfo{pages}{6445--6459}.
\newblock


\bibitem[Garc{\'\i}a et~al\mbox{.}(2015)]%
        {garcia2015robust}
\bibfield{author}{\bibinfo{person}{Enrique Garc{\'\i}a}, \bibinfo{person}{Pablo
  Poudereux}, \bibinfo{person}{{\'A}lvaro Hern{\'a}ndez},
  \bibinfo{person}{Jes{\'u}s Ure{\~n}a}, {and} \bibinfo{person}{David Gualda}.}
  \bibinfo{year}{2015}\natexlab{}.
\newblock \showarticletitle{A robust UWB indoor positioning system for highly
  complex environments}. In \bibinfo{booktitle}{\emph{IEEE International
  Conference on Industrial Technology (ICIT)}}. \bibinfo{pages}{3386--3391}.
\newblock


\bibitem[Ge and Shen(2021)]%
        {ge2021single}
\bibfield{author}{\bibinfo{person}{Feng Ge} {and} \bibinfo{person}{Yuan Shen}.}
  \bibinfo{year}{2021}\natexlab{}.
\newblock \showarticletitle{Single-anchor ultra-wideband localization system
  using wrapped PDoA}.
\newblock \bibinfo{journal}{\emph{IEEE Transactions on Mobile Computing}}
  \bibinfo{volume}{21}, \bibinfo{number}{12} (\bibinfo{year}{2021}),
  \bibinfo{pages}{4609--4623}.
\newblock


\bibitem[Gil et~al\mbox{.}(2007)]%
        {gil2007numerical}
\bibfield{author}{\bibinfo{person}{Amparo Gil}, \bibinfo{person}{Javier
  Segura}, {and} \bibinfo{person}{Nico Temme}.}
  \bibinfo{year}{2007}\natexlab{}.
\newblock \bibinfo{booktitle}{\emph{Numerical methods for special functions}}.
\newblock \bibinfo{publisher}{SIAM}.
\newblock


\bibitem[Giorgetti et~al\mbox{.}(2009)]%
        {giorgetti2009single}
\bibfield{author}{\bibinfo{person}{Gianni Giorgetti},
  \bibinfo{person}{Alessandro Cidronali}, \bibinfo{person}{Sandeep Gupta},
  {and} \bibinfo{person}{Gianfranco Manes}.} \bibinfo{year}{2009}\natexlab{}.
\newblock \showarticletitle{Single-anchor indoor localization using a
  switched-beam antenna}.
\newblock \bibinfo{journal}{\emph{IEEE Communications Letters}}
  \bibinfo{volume}{13}, \bibinfo{number}{1} (\bibinfo{year}{2009}),
  \bibinfo{pages}{58--60}.
\newblock


\bibitem[Glorot and Bengio(2010)]%
        {glorot2010understanding}
\bibfield{author}{\bibinfo{person}{Xavier Glorot} {and} \bibinfo{person}{Yoshua
  Bengio}.} \bibinfo{year}{2010}\natexlab{}.
\newblock \showarticletitle{Understanding the difficulty of training deep
  feedforward neural networks}. In \bibinfo{booktitle}{\emph{International
  Conference on Artificial Intelligence and Statistics (AISTATS)}}.
  \bibinfo{pages}{249--256}.
\newblock


\bibitem[Gowda et~al\mbox{.}(2017)]%
        {iball}
\bibfield{author}{\bibinfo{person}{Mahanth Gowda}, \bibinfo{person}{Ashutosh
  Dhekne}, \bibinfo{person}{Sheng Shen}, \bibinfo{person}{Romit~Roy Choudhury},
  \bibinfo{person}{Lei Yang}, \bibinfo{person}{Suresh Golwalkar}, {and}
  \bibinfo{person}{Alexander Essanian}.} \bibinfo{year}{2017}\natexlab{}.
\newblock \showarticletitle{Bringing {IoT} to Sports Analytics}. In
  \bibinfo{booktitle}{\emph{{USENIX} Symposium on Networked Systems Design and
  Implementation ({NSDI})}}. \bibinfo{pages}{499--513}.
\newblock


\bibitem[Gro$\beta$windhager et~al\mbox{.}(2019)]%
        {grobetawindhager2019snaploc}
\bibfield{author}{\bibinfo{person}{Bernhard Gro$\beta$windhager},
  \bibinfo{person}{Michael Stocker}, \bibinfo{person}{Michael Rath},
  \bibinfo{person}{Carlo~Alberto Boano}, {and} \bibinfo{person}{Kay
  R{\"o}mer}.} \bibinfo{year}{2019}\natexlab{}.
\newblock \showarticletitle{SnapLoc: An ultra-fast UWB-based indoor
  localization system for an unlimited number of tags}. In
  \bibinfo{booktitle}{\emph{ACM/IEEE International Conference on Information
  Processing in Sensor Networks (IPSN)}}. \bibinfo{pages}{61--72}.
\newblock


\bibitem[Gro{\ss}windhager et~al\mbox{.}(2018)]%
        {grosswindhager2018salma}
\bibfield{author}{\bibinfo{person}{Bernhard Gro{\ss}windhager},
  \bibinfo{person}{Michael Rath}, \bibinfo{person}{Josef Kulmer},
  \bibinfo{person}{Mustafa~S Bakr}, \bibinfo{person}{Carlo~Alberto Boano},
  \bibinfo{person}{Klaus Witrisal}, {and} \bibinfo{person}{Kay R{\"o}mer}.}
  \bibinfo{year}{2018}\natexlab{}.
\newblock \showarticletitle{SALMA: UWB-based single-anchor localization system
  using multipath assistance}. In \bibinfo{booktitle}{\emph{ACM Conference on
  Embedded Networked Sensor Systems (SenSys)}}. \bibinfo{pages}{132--144}.
\newblock


\bibitem[Groth et~al\mbox{.}(2021)]%
        {groth2021calibration}
\bibfield{author}{\bibinfo{person}{Mateusz Groth}, \bibinfo{person}{Krzysztof
  Nyka}, {and} \bibinfo{person}{Lukasz Kulas}.}
  \bibinfo{year}{2021}\natexlab{}.
\newblock \showarticletitle{Calibration-free single-anchor indoor localization
  using an ESPAR antenna}.
\newblock \bibinfo{journal}{\emph{Sensors}} \bibinfo{volume}{21},
  \bibinfo{number}{10} (\bibinfo{year}{2021}), \bibinfo{pages}{3431:1--21}.
\newblock


\bibitem[Gu et~al\mbox{.}(2021)]%
        {gu2021tyrloc}
\bibfield{author}{\bibinfo{person}{Zhihao Gu}, \bibinfo{person}{Taiwei He},
  \bibinfo{person}{Junwei Yin}, \bibinfo{person}{Yuedong Xu}, {and}
  \bibinfo{person}{Jun Wu}.} \bibinfo{year}{2021}\natexlab{}.
\newblock \showarticletitle{TyrLoc: A low-cost multi-technology MIMO
  localization system with a single RF chain}. In \bibinfo{booktitle}{\emph{ACM
  International Conference on Mobile Systems, Applications, and Services
  (MobiSys)}}. \bibinfo{pages}{228--240}.
\newblock


\bibitem[Guo et~al\mbox{.}(2022)]%
        {guo2022multi}
\bibfield{author}{\bibinfo{person}{Yundong Guo}, \bibinfo{person}{Zhenyu Liu},
  \bibinfo{person}{Hao Luo}, \bibinfo{person}{Huijie Pu}, {and}
  \bibinfo{person}{Jianrong Tan}.} \bibinfo{year}{2022}\natexlab{}.
\newblock \showarticletitle{Multi-person multi-camera tracking for live stream
  videos based on improved motion model and matching cascade}.
\newblock \bibinfo{journal}{\emph{Neurocomputing}}  \bibinfo{volume}{492}
  (\bibinfo{year}{2022}), \bibinfo{pages}{561--571}.
\newblock


\bibitem[Gupta(2016)]%
        {gupta2016ble}
\bibfield{author}{\bibinfo{person}{Naresh~Kumar Gupta}.}
  \bibinfo{year}{2016}\natexlab{}.
\newblock \bibinfo{booktitle}{\emph{Inside Bluetooth low energy}}.
\newblock \bibinfo{publisher}{Artech House}.
\newblock


\bibitem[Hasan et~al\mbox{.}(2022)]%
        {hasan2022lidar}
\bibfield{author}{\bibinfo{person}{Mahmudul Hasan}, \bibinfo{person}{Junichi
  Hanawa}, \bibinfo{person}{Riku Goto}, \bibinfo{person}{Ryota Suzuki},
  \bibinfo{person}{Hisato Fukuda}, \bibinfo{person}{Yoshinori Kuno}, {and}
  \bibinfo{person}{Yoshinori Kobayashi}.} \bibinfo{year}{2022}\natexlab{}.
\newblock \showarticletitle{LiDAR-based detection, tracking, and property
  estimation: A contemporary review}.
\newblock \bibinfo{journal}{\emph{Neurocomputing}}  \bibinfo{volume}{206}
  (\bibinfo{year}{2022}), \bibinfo{pages}{393--405}.
\newblock


\bibitem[Heydariaan et~al\mbox{.}(2020)]%
        {heydariaan2020anguloc}
\bibfield{author}{\bibinfo{person}{Milad Heydariaan}, \bibinfo{person}{Hossein
  Dabirian}, {and} \bibinfo{person}{Omprakash Gnawali}.}
  \bibinfo{year}{2020}\natexlab{}.
\newblock \showarticletitle{AnguLoc: Concurrent angle of arrival estimation for
  indoor localization with UWB radios}. In \bibinfo{booktitle}{\emph{IEEE
  International Conference on Distributed Computing in Sensor Systems
  (DCOSS)}}. \bibinfo{pages}{112--119}.
\newblock


\bibitem[Islam et~al\mbox{.}(2019)]%
        {percom2019lorain}
\bibfield{author}{\bibinfo{person}{Bashima Islam}, \bibinfo{person}{Md~Tamzeed
  Islam}, \bibinfo{person}{Jasleen Kaur}, {and} \bibinfo{person}{Shahriar
  Nirjon}.} \bibinfo{year}{2019}\natexlab{}.
\newblock \showarticletitle{Lorain: Making a case for lora in indoor
  localization}. In \bibinfo{booktitle}{\emph{IEEE International Conference on
  Pervasive Computing and Communications Workshops (PerCom Workshops)}}.
  \bibinfo{pages}{423--426}.
\newblock


\bibitem[Jiang et~al\mbox{.}(2018)]%
        {ipsn2018omnitrack}
\bibfield{author}{\bibinfo{person}{Chengkun Jiang}, \bibinfo{person}{Yuan He},
  \bibinfo{person}{Xiaolong Zheng}, {and} \bibinfo{person}{Yunhao Liu}.}
  \bibinfo{year}{2018}\natexlab{}.
\newblock \showarticletitle{Orientation-aware RFID tracking with
  centimeter-level accuracy}. In \bibinfo{booktitle}{\emph{ACM/IEEE
  International Conference on Information Processing in Sensor Networks
  (IPSN)}}. \bibinfo{pages}{290--301}.
\newblock


\bibitem[Kempke et~al\mbox{.}(2016)]%
        {kempke2016harmonium}
\bibfield{author}{\bibinfo{person}{Benjamin Kempke}, \bibinfo{person}{Pat
  Pannuto}, {and} \bibinfo{person}{Prabal Dutta}.}
  \bibinfo{year}{2016}\natexlab{}.
\newblock \showarticletitle{Harmonium: Asymmetric, bandstitched UWB for fast,
  accurate, and robust indoor localization}. In
  \bibinfo{booktitle}{\emph{ACM/IEEE International Conference on Information
  Processing in Sensor Networks (IPSN)}}. \bibinfo{pages}{1--12}.
\newblock


\bibitem[Kong et~al\mbox{.}(2022)]%
        {kong2022m3track}
\bibfield{author}{\bibinfo{person}{Hao Kong}, \bibinfo{person}{Xiangyu Xu},
  \bibinfo{person}{Jiadi Yu}, \bibinfo{person}{Qilin Chen},
  \bibinfo{person}{Chenguang Ma}, \bibinfo{person}{Yingying Chen},
  \bibinfo{person}{Yi-Chao Chen}, {and} \bibinfo{person}{Linghe Kong}.}
  \bibinfo{year}{2022}\natexlab{}.
\newblock \showarticletitle{m$^3$Track: mmWave-based multi-user 3D posture
  tracking}. In \bibinfo{booktitle}{\emph{ACM International Conference on
  Mobile Systems, Applications, and Services (MobiSys)}}.
  \bibinfo{pages}{491--503}.
\newblock


\bibitem[Kotaru et~al\mbox{.}(2015)]%
        {spotfi}
\bibfield{author}{\bibinfo{person}{Manikanta Kotaru}, \bibinfo{person}{Kiran
  Joshi}, \bibinfo{person}{Dinesh Bharadia}, {and} \bibinfo{person}{Sachin
  Katti}.} \bibinfo{year}{2015}\natexlab{}.
\newblock \showarticletitle{{SpotFi: Decimeter level localization using
  Wi-Fi}}. In \bibinfo{booktitle}{\emph{ACM Conference on Special Interest
  Group on Data Communication (SIGCOMM)}}. \bibinfo{pages}{269--282}.
\newblock


\bibitem[Kotaru and Katti(2017)]%
        {kotaru2017position}
\bibfield{author}{\bibinfo{person}{Manikanta Kotaru} {and}
  \bibinfo{person}{Sachin Katti}.} \bibinfo{year}{2017}\natexlab{}.
\newblock \showarticletitle{Position tracking for virtual reality using
  commodity WiFi}. In \bibinfo{booktitle}{\emph{IEEE Conference on Computer
  Vision and Pattern Recognition (CVPR)}}. \bibinfo{pages}{68--78}.
\newblock


\bibitem[Ledergerber and D’Andrea(2019)]%
        {ledergerber2019ultra}
\bibfield{author}{\bibinfo{person}{Anton Ledergerber} {and}
  \bibinfo{person}{Raffaello D’Andrea}.} \bibinfo{year}{2019}\natexlab{}.
\newblock \showarticletitle{Ultra-wideband angle of arrival estimation based on
  angle-dependent antenna transfer function}.
\newblock \bibinfo{journal}{\emph{Sensors}} \bibinfo{volume}{19},
  \bibinfo{number}{20} (\bibinfo{year}{2019}), \bibinfo{pages}{4466:1--21}.
\newblock


\bibitem[Li et~al\mbox{.}(2022)]%
        {li2022experience}
\bibfield{author}{\bibinfo{person}{Dong Li}, \bibinfo{person}{Shirui Cao},
  \bibinfo{person}{Sunghoon~Ivan Lee}, {and} \bibinfo{person}{Jie Xiong}.}
  \bibinfo{year}{2022}\natexlab{}.
\newblock \showarticletitle{Experience: Practical problems for acoustic
  sensing}. In \bibinfo{booktitle}{\emph{ACM Annual International Conference on
  Mobile Computing and Networking (MobiCom)}}. \bibinfo{pages}{381--390}.
\newblock


\bibitem[Li et~al\mbox{.}(2020a)]%
        {sensys2020fmtrack}
\bibfield{author}{\bibinfo{person}{Dong Li}, \bibinfo{person}{Jialin Liu},
  \bibinfo{person}{Sunghoon~Ivan Lee}, {and} \bibinfo{person}{Jie Xiong}.}
  \bibinfo{year}{2020}\natexlab{a}.
\newblock \showarticletitle{FM-Track: Pushing the limits of contactless
  multi-target tracking using acoustic signals}. In
  \bibinfo{booktitle}{\emph{ACM Conference on Embedded Networked Sensor Systems
  (SenSys)}}. \bibinfo{pages}{150--163}.
\newblock


\bibitem[Li et~al\mbox{.}(2020c)]%
        {li2020pose}
\bibfield{author}{\bibinfo{person}{Jing Li}, \bibinfo{person}{Jing Xu},
  \bibinfo{person}{Fangwei Zhong}, \bibinfo{person}{Xiangyu Kong},
  \bibinfo{person}{Yu Qiao}, {and} \bibinfo{person}{Yizhou Wang}.}
  \bibinfo{year}{2020}\natexlab{c}.
\newblock \showarticletitle{Pose-assisted multi-camera collaboration for active
  object tracking}. In \bibinfo{booktitle}{\emph{AAAI Conference on Artificial
  Intelligence (AAAI)}}, Vol.~\bibinfo{volume}{34}. \bibinfo{pages}{759--766}.
\newblock


\bibitem[Li et~al\mbox{.}(2020b)]%
        {li2020multipath}
\bibfield{author}{\bibinfo{person}{Ze Li}, \bibinfo{person}{Zengshan Tian},
  \bibinfo{person}{Zhongchun Wang}, {and} \bibinfo{person}{Zhenyuan Zhang}.}
  \bibinfo{year}{2020}\natexlab{b}.
\newblock \showarticletitle{Multipath-assisted indoor localization using a
  single receiver}.
\newblock \bibinfo{journal}{\emph{IEEE Sensors Journal}} \bibinfo{volume}{21},
  \bibinfo{number}{1} (\bibinfo{year}{2020}), \bibinfo{pages}{692--705}.
\newblock


\bibitem[Liu et~al\mbox{.}(2013)]%
        {liu2013joint}
\bibfield{author}{\bibinfo{person}{Congfeng Liu}, \bibinfo{person}{Jie Yang},
  {and} \bibinfo{person}{Fengshuai Wang}.} \bibinfo{year}{2013}\natexlab{}.
\newblock \showarticletitle{Joint TDoA and AoA location algorithm}.
\newblock \bibinfo{journal}{\emph{Journal of Systems Engineering and
  Electronics}} \bibinfo{volume}{24}, \bibinfo{number}{2}
  (\bibinfo{year}{2013}), \bibinfo{pages}{183--188}.
\newblock


\bibitem[Liu et~al\mbox{.}(2020)]%
        {liu2020indoor}
\bibfield{author}{\bibinfo{person}{Manni Liu}, \bibinfo{person}{Linsong Cheng},
  \bibinfo{person}{Kun Qian}, \bibinfo{person}{Jiliang Wang},
  \bibinfo{person}{Jin Wang}, {and} \bibinfo{person}{Yunhao Liu}.}
  \bibinfo{year}{2020}\natexlab{}.
\newblock \showarticletitle{Indoor acoustic localization: A survey}.
\newblock \bibinfo{journal}{\emph{Human-centric Computing and Information
  Sciences}}  \bibinfo{volume}{10} (\bibinfo{year}{2020}),
  \bibinfo{pages}{1--24}.
\newblock


\bibitem[Luo et~al\mbox{.}(2022)]%
        {luo2022novel}
\bibfield{author}{\bibinfo{person}{Donghai Luo}, \bibinfo{person}{Daobo Wang},
  \bibinfo{person}{Shengji Xia}, {and} \bibinfo{person}{Tingting Bai}.}
  \bibinfo{year}{2022}\natexlab{}.
\newblock \showarticletitle{A novel approach for visual tracking based on
  occlusion recognition}.
\newblock \bibinfo{journal}{\emph{Highlights in Science, Engineering and
  Technology}}  \bibinfo{volume}{7} (\bibinfo{year}{2022}),
  \bibinfo{pages}{124--133}.
\newblock


\bibitem[Luo et~al\mbox{.}(2019)]%
        {turbotrack}
\bibfield{author}{\bibinfo{person}{Zhihong Luo}, \bibinfo{person}{Qiping
  Zhang}, \bibinfo{person}{Yunfei Ma}, \bibinfo{person}{Manish Singh}, {and}
  \bibinfo{person}{Fadel Adib}.} \bibinfo{year}{2019}\natexlab{}.
\newblock \showarticletitle{{3D} backscatter localization for fine-grained
  robotics}. In \bibinfo{booktitle}{\emph{{USENIX} Symposium on Networked
  Systems Design and Implementation ({NSDI})}}. \bibinfo{pages}{765--782}.
\newblock


\bibitem[Ma et~al\mbox{.}(2017)]%
        {ma2017minding}
\bibfield{author}{\bibinfo{person}{Yunfei Ma}, \bibinfo{person}{Nicholas
  Selby}, {and} \bibinfo{person}{Fadel Adib}.} \bibinfo{year}{2017}\natexlab{}.
\newblock \showarticletitle{Minding the billions: Ultra-wideband localization
  for deployed {RFID} tags}. In \bibinfo{booktitle}{\emph{ACM Annual
  International Conference on Mobile Computing and Networking (MobiCom)}}.
  \bibinfo{pages}{248--260}.
\newblock


\bibitem[Macoir et~al\mbox{.}(2018)]%
        {macoir2018mac}
\bibfield{author}{\bibinfo{person}{Nicola Macoir}, \bibinfo{person}{Matteo
  Ridolfi}, \bibinfo{person}{Jen Rossey}, \bibinfo{person}{Ingrid Moerman},
  {and} \bibinfo{person}{Eli De~Poorter}.} \bibinfo{year}{2018}\natexlab{}.
\newblock \showarticletitle{MAC protocol for supporting multiple roaming users
  in mult-cell UWB localization networks}. In \bibinfo{booktitle}{\emph{IEEE
  International Symposium on A World of Wireless, Mobile and Multimedia
  Networks (WoWMoM)}}. \bibinfo{pages}{588--599}.
\newblock


\bibitem[Meissner and Witrisal(2012)]%
        {meissner2012multipath}
\bibfield{author}{\bibinfo{person}{Paul Meissner} {and} \bibinfo{person}{Klaus
  Witrisal}.} \bibinfo{year}{2012}\natexlab{}.
\newblock \showarticletitle{Multipath-assisted single-anchor indoor
  localization in an office environment}. In
  \bibinfo{booktitle}{\emph{International Conference on Systems, Signals and
  Image Processing (IWSSIP)}}. \bibinfo{pages}{22--25}.
\newblock


\bibitem[Merenda et~al\mbox{.}(2022)]%
        {merenda2022rfid}
\bibfield{author}{\bibinfo{person}{Massimo Merenda}, \bibinfo{person}{Luca
  Catarinucci}, \bibinfo{person}{Riccardo Colella}, \bibinfo{person}{Demetrio
  Iero}, \bibinfo{person}{Francesco Della~Corte}, {and}
  \bibinfo{person}{Riccardo Carotenuto}.} \bibinfo{year}{2022}\natexlab{}.
\newblock \showarticletitle{RFID-based indoor positioning using edge machine
  learning}.
\newblock \bibinfo{journal}{\emph{IEEE Journal of Radio Frequency
  Identification}}  \bibinfo{volume}{6} (\bibinfo{year}{2022}),
  \bibinfo{pages}{573--582}.
\newblock


\bibitem[Miller et~al\mbox{.}(2022)]%
        {miller2022cappella}
\bibfield{author}{\bibinfo{person}{John Miller}, \bibinfo{person}{Elahe
  Soltanaghai}, \bibinfo{person}{Raewyn Duvall}, \bibinfo{person}{Jeff Chen},
  \bibinfo{person}{Vikram Bhat}, \bibinfo{person}{Nuno Pereira}, {and}
  \bibinfo{person}{Anthony Rowe}.} \bibinfo{year}{2022}\natexlab{}.
\newblock \showarticletitle{Cappella: Establishing multi-user augmented reality
  sessions using inertial estimates and peer-to-peer ranging}. In
  \bibinfo{booktitle}{\emph{ACM/IEEE International Conference on Information
  Processing in Sensor Networks (IPSN)}}. \bibinfo{pages}{428--440}.
\newblock


\bibitem[Monica and Aleotti(2022)]%
        {monica2022evaluation}
\bibfield{author}{\bibinfo{person}{Riccardo Monica} {and}
  \bibinfo{person}{Jacopo Aleotti}.} \bibinfo{year}{2022}\natexlab{}.
\newblock \showarticletitle{Evaluation of the Oculus Rift S tracking system in
  room scale virtual reality}.
\newblock \bibinfo{journal}{\emph{Virtual Reality}} \bibinfo{volume}{26},
  \bibinfo{number}{4} (\bibinfo{year}{2022}), \bibinfo{pages}{1335--1345}.
\newblock


\bibitem[Mostafa et~al\mbox{.}(2022)]%
        {mostafa2022survey}
\bibfield{author}{\bibinfo{person}{Sherif Mostafa}, \bibinfo{person}{Khaled
  Harras}, {and} \bibinfo{person}{Moustafa Youssef}.}
  \bibinfo{year}{2022}\natexlab{}.
\newblock \showarticletitle{A survey of indoor localization systems in
  multi-floor environments}.
\newblock \bibinfo{journal}{\emph{TechRxiv}} (\bibinfo{year}{2022}),
  \bibinfo{pages}{1--31}.
\newblock


\bibitem[Mukherjee et~al\mbox{.}(2022)]%
        {mukherjee2022scalable}
\bibfield{author}{\bibinfo{person}{Swagato Mukherjee}, \bibinfo{person}{Gregory
  Skidmore}, \bibinfo{person}{Tarun Chawla}, \bibinfo{person}{Anmol Bhardwaj},
  \bibinfo{person}{Camillo Gentile}, {and} \bibinfo{person}{Jelena Senic}.}
  \bibinfo{year}{2022}\natexlab{}.
\newblock \showarticletitle{Scalable modeling of human blockage at
  millimeter-wave: A comparative analysis of knife-edge diffraction, the
  uniform theory of diffraction, and physical optics against 60 GHz channel
  measurements}.
\newblock \bibinfo{journal}{\emph{IEEE Access}}  \bibinfo{volume}{10}
  (\bibinfo{year}{2022}), \bibinfo{pages}{133643--133654}.
\newblock


\bibitem[Murakami et~al\mbox{.}(2019)]%
        {murakami20193}
\bibfield{author}{\bibinfo{person}{Hiroaki Murakami}, \bibinfo{person}{Masanari
  Nakamura}, \bibinfo{person}{Hiromichi Hashizume}, {and}
  \bibinfo{person}{Masanori Sugimoto}.} \bibinfo{year}{2019}\natexlab{}.
\newblock \showarticletitle{3-D localization for smartphones using a single
  speaker}. In \bibinfo{booktitle}{\emph{International Conference on Indoor
  Positioning and Indoor Navigation (IPIN)}}. \bibinfo{pages}{1--8}.
\newblock


\bibitem[{Poulose} et~al\mbox{.}(2020)]%
        {9289338}
\bibfield{author}{\bibinfo{person}{A. {Poulose}}, \bibinfo{person}{Z.
  {Emersic}}, \bibinfo{person}{O. {Steven Eyobu}}, {and} \bibinfo{person}{D.
  {Seog Han}}.} \bibinfo{year}{2020}\natexlab{}.
\newblock \showarticletitle{An accurate indoor user position estimator for
  multiple anchor UWB localization}. In \bibinfo{booktitle}{\emph{International
  Conference on Information and Communication Technology Convergence (ICTC)}}.
  \bibinfo{pages}{478--482}.
\newblock


\bibitem[{Qorvo, Inc.}(2016)]%
        {evb1000}
\bibfield{author}{\bibinfo{person}{{Qorvo, Inc.}}}
  \bibinfo{year}{2016}\natexlab{}.
\newblock \bibinfo{title}{EVB1000: Ultra-Wideband (UWB) Transceiver Evaluation
  Kit}.
\newblock
  \bibinfo{howpublished}{\url{https://www.qorvo.com/products/p/EVK1000}}.
\newblock
\newblock
\shownote{Accessed: 2023-02-14}.


\bibitem[{Qurvo, Inc.}(2014)]%
        {tdoa_an}
\bibfield{author}{\bibinfo{person}{{Qurvo, Inc.}}}
  \bibinfo{year}{2014}\natexlab{}.
\newblock \bibinfo{title}{Wired Synchronization of Anchor Nodes in a TDoA Real
  Time Location System}.
\newblock
\newblock
\newblock
\shownote{APS007, version 1.10}.


\bibitem[{Qurvo, Inc.}(2016)]%
        {dw_fpi}
\bibfield{author}{\bibinfo{person}{{Qurvo, Inc.}}}
  \bibinfo{year}{2016}\natexlab{}.
\newblock \bibinfo{title}{DW1000 Metrics for Estimation of Non Line Of Sight
  Operating Conditions}.
\newblock
\newblock
\newblock
\shownote{APS006, version 1.10}.


\bibitem[Ramirez et~al\mbox{.}(2019)]%
        {ramirez2019longshot}
\bibfield{author}{\bibinfo{person}{Ceferino~Gabriel Ramirez},
  \bibinfo{person}{Anton Sergeyev}, \bibinfo{person}{Assya Dyussenova}, {and}
  \bibinfo{person}{Bob Iannucci}.} \bibinfo{year}{2019}\natexlab{}.
\newblock \showarticletitle{LongShoT: Long-range synchronization of time}. In
  \bibinfo{booktitle}{\emph{ACM/IEEE International Conference on Information
  Processing in Sensor Networks (IPSN)}}. \bibinfo{pages}{289--300}.
\newblock


\bibitem[Rostami and Sundaresan(2022)]%
        {rostami2022enabling}
\bibfield{author}{\bibinfo{person}{Mohammad Rostami} {and}
  \bibinfo{person}{Karthikeyan Sundaresan}.} \bibinfo{year}{2022}\natexlab{}.
\newblock \showarticletitle{Enabling high accuracy pervasive tracking with
  ultra low power UWB tags}. In \bibinfo{booktitle}{\emph{ACM Annual
  International Conference on Mobile Computing and Networking (MobiCom)}}.
  \bibinfo{pages}{459--472}.
\newblock


\bibitem[Russell(2013)]%
        {russell2013people}
\bibfield{author}{\bibinfo{person}{Kyle Russell}.}
  \bibinfo{year}{2013}\natexlab{}.
\newblock \showarticletitle{People are worried Microsoft’s new Xbox will be
  able to spy on you}.
\newblock \bibinfo{journal}{\emph{Business Insider}} (\bibinfo{year}{2013}).
\newblock


\bibitem[Sahin and Itti(2023)]%
        {sahin2023hoot}
\bibfield{author}{\bibinfo{person}{Gozde Sahin} {and} \bibinfo{person}{Laurent
  Itti}.} \bibinfo{year}{2023}\natexlab{}.
\newblock \showarticletitle{HOOT: Heavy occlusions in object tracking
  benchmark}. In \bibinfo{booktitle}{\emph{IEEE/CVF Winter Conference on
  Applications of Computer Vision (WACV)}}. \bibinfo{pages}{4830--4839}.
\newblock


\bibitem[Sanchez-Iborra and Cano(2016)]%
        {sanchez2016state}
\bibfield{author}{\bibinfo{person}{Ramon Sanchez-Iborra} {and}
  \bibinfo{person}{Maria-Dolores Cano}.} \bibinfo{year}{2016}\natexlab{}.
\newblock \showarticletitle{State of the art in LP-WAN solutions for industrial
  IoT services}.
\newblock \bibinfo{journal}{\emph{Sensors}} \bibinfo{volume}{16},
  \bibinfo{number}{5} (\bibinfo{year}{2016}), \bibinfo{pages}{708:1--14}.
\newblock


\bibitem[Scheuing and Yang(2006)]%
        {scheuing2006disambiguation}
\bibfield{author}{\bibinfo{person}{Jan Scheuing} {and} \bibinfo{person}{Bin
  Yang}.} \bibinfo{year}{2006}\natexlab{}.
\newblock \showarticletitle{Disambiguation of TDoA estimates in multi-path
  multi-source environments (DATEMM)}. In \bibinfo{booktitle}{\emph{IEEE
  International Conference on Acoustics Speech and Signal Processing
  Proceedings (ICASSP)}}, Vol.~\bibinfo{volume}{4}. \bibinfo{pages}{837--840}.
\newblock


\bibitem[{Semtech Corp.}(2022)]%
        {sx1272mb2das}
\bibfield{author}{\bibinfo{person}{{Semtech Corp.}}}
  \bibinfo{year}{2022}\natexlab{}.
\newblock \bibinfo{title}{LoRa Connect Mbed shield, SX1272, 868 and 915MHz}.
\newblock
  \bibinfo{howpublished}{\url{https://www.semtech.com/products/wireless-rf/lora-connect/sx1272mb2das}}.
\newblock
\newblock
\shownote{Accessed: 2023-03-17}.


\bibitem[Shangguan and Jamieson(2016)]%
        {mobitagbot}
\bibfield{author}{\bibinfo{person}{Longfei Shangguan} {and}
  \bibinfo{person}{Kyle Jamieson}.} \bibinfo{year}{2016}\natexlab{}.
\newblock \showarticletitle{The design and implementation of a mobile {RFID}
  tag sorting robot}. In \bibinfo{booktitle}{\emph{ACM International Conference
  on Mobile Systems, Applications, and Services (MobiSys)}}.
  \bibinfo{pages}{31--42}.
\newblock


\bibitem[Slezak and Rangan(2022)]%
        {slezak2022measurement}
\bibfield{author}{\bibinfo{person}{Christopher Slezak} {and}
  \bibinfo{person}{Sundeep Rangan}.} \bibinfo{year}{2022}\natexlab{}.
\newblock \showarticletitle{Measurement-based indoor millimeter wave blockage
  models}.
\newblock \bibinfo{journal}{\emph{IEEE Transactions on Wireless
  Communications}} \bibinfo{volume}{21}, \bibinfo{number}{8}
  (\bibinfo{year}{2022}), \bibinfo{pages}{6774--6786}.
\newblock


\bibitem[Soltanaghaei et~al\mbox{.}(2018)]%
        {soltanaghaei2018multipath}
\bibfield{author}{\bibinfo{person}{Elahe Soltanaghaei},
  \bibinfo{person}{Avinash Kalyanaraman}, {and} \bibinfo{person}{Kamin
  Whitehouse}.} \bibinfo{year}{2018}\natexlab{}.
\newblock \showarticletitle{Multipath triangulation: Decimeter-level WiFi
  localization and orientation with a single unaided receiver}. In
  \bibinfo{booktitle}{\emph{ACM International Conference on Mobile Systems,
  Applications, and Services (MobiSys)}}. \bibinfo{pages}{376--378}.
\newblock


\bibitem[Soltanaghaei et~al\mbox{.}(2021)]%
        {soltanaghaei2021millimetro}
\bibfield{author}{\bibinfo{person}{Elahe Soltanaghaei}, \bibinfo{person}{Akarsh
  Prabhakara}, \bibinfo{person}{Artur Balanuta}, \bibinfo{person}{Matthew
  Anderson}, \bibinfo{person}{Jan Rabaey}, \bibinfo{person}{Swarun Kumar},
  {and} \bibinfo{person}{Anthony Rowe}.} \bibinfo{year}{2021}\natexlab{}.
\newblock \showarticletitle{Millimetro: mmWave retro-reflective tags for
  accurate, long range localization}. In \bibinfo{booktitle}{\emph{ACM Annual
  International Conference on Mobile Computing and Networking (MobiCom)}}.
  \bibinfo{pages}{69--82}.
\newblock


\bibitem[Spilker~Jr et~al\mbox{.}(1996)]%
        {spilker1996global}
\bibfield{author}{\bibinfo{person}{James Spilker~Jr}, \bibinfo{person}{Penina
  Axelrad}, \bibinfo{person}{Bradford Parkinson}, {and} \bibinfo{person}{Per
  Enge}.} \bibinfo{year}{1996}\natexlab{}.
\newblock \bibinfo{booktitle}{\emph{Global positioning system: Theory and
  applications, volume I}}.
\newblock \bibinfo{publisher}{American Institute of Aeronautics and
  Astronautics}.
\newblock


\bibitem[{Texas Instruments, Inc.}(2017)]%
        {lmk-clock}
\bibfield{author}{\bibinfo{person}{{Texas Instruments, Inc.}}}
  \bibinfo{year}{2017}\natexlab{}.
\newblock \bibinfo{title}{LMK04832EVM: LMK04832 evaluation module for
  ultra-low-noise, 3.2-GHz, 15-output, JESD204B clock jitter cleaner}.
\newblock \bibinfo{howpublished}{\url{https://www.ti.com/tool/LMK04832EVM}}.
\newblock
\newblock
\shownote{Accessed: 2023-02-14}.


\bibitem[Thomas(2012)]%
        {thomas2012survey}
\bibfield{author}{\bibinfo{person}{Bruce Thomas}.}
  \bibinfo{year}{2012}\natexlab{}.
\newblock \showarticletitle{A survey of visual, mixed, and augmented reality
  gaming}.
\newblock \bibinfo{journal}{\emph{Computers in Entertainment}}
  \bibinfo{volume}{10}, \bibinfo{number}{1} (\bibinfo{year}{2012}),
  \bibinfo{pages}{1--33}.
\newblock


\bibitem[Tiemann et~al\mbox{.}(2016)]%
        {tiemann2016atlas}
\bibfield{author}{\bibinfo{person}{Janis Tiemann}, \bibinfo{person}{Fabian
  Eckermann}, {and} \bibinfo{person}{Christian Wietfeld}.}
  \bibinfo{year}{2016}\natexlab{}.
\newblock \showarticletitle{Atlas: an open-source tdoa-based ultra-wideband
  localization system}. In \bibinfo{booktitle}{\emph{International Conference
  on Indoor Positioning and Indoor Navigation (IPIN)}}. \bibinfo{pages}{1--6}.
\newblock


\bibitem[Tiemann et~al\mbox{.}(2019)]%
        {tiemann2019atlas}
\bibfield{author}{\bibinfo{person}{Janis Tiemann}, \bibinfo{person}{Yehya
  Elmasry}, \bibinfo{person}{Lucas Koring}, {and} \bibinfo{person}{Christian
  Wietfeld}.} \bibinfo{year}{2019}\natexlab{}.
\newblock \showarticletitle{ATLAS FaST: Fast and simple scheduled TDOA for
  reliable ultra-wideband localization}. In \bibinfo{booktitle}{\emph{IEEE
  International Conference on Robotics and Automation (ICRA)}}.
  \bibinfo{pages}{2554--2560}.
\newblock


\bibitem[Vaidyanathan and Pal(2010)]%
        {vaidyanathan2010sparse}
\bibfield{author}{\bibinfo{person}{P. Vaidyanathan} {and} \bibinfo{person}{Piya
  Pal}.} \bibinfo{year}{2010}\natexlab{}.
\newblock \showarticletitle{Sparse sensing with coprime arrays}. In
  \bibinfo{booktitle}{\emph{Asilomar Conference on Signals, Systems and
  Computers}}. \bibinfo{pages}{1405--1409}.
\newblock


\bibitem[Vasisht et~al\mbox{.}(2016)]%
        {vasisht2016decimeter}
\bibfield{author}{\bibinfo{person}{Deepak Vasisht}, \bibinfo{person}{Swarun
  Kumar}, {and} \bibinfo{person}{Dina Katabi}.}
  \bibinfo{year}{2016}\natexlab{}.
\newblock \showarticletitle{Decimeter-level localization with a single WiFi
  access point}. In \bibinfo{booktitle}{\emph{{USENIX} Symposium on Networked
  Systems Design and Implementation ({NSDI})}}. \bibinfo{pages}{165--178}.
\newblock


\bibitem[Vecchia et~al\mbox{.}(2019)]%
        {vecchia2019talla}
\bibfield{author}{\bibinfo{person}{Davide Vecchia}, \bibinfo{person}{Pablo
  Corbal{\'a}n}, \bibinfo{person}{Timofei Istomin}, {and}
  \bibinfo{person}{Gian~Pietro Picco}.} \bibinfo{year}{2019}\natexlab{}.
\newblock \showarticletitle{TALLA: Large-scale TDoA localization with
  ultra-wideband radios}. In \bibinfo{booktitle}{\emph{International Conference
  on Indoor Positioning and Indoor Navigation (IPIN)}}. \bibinfo{pages}{1--8}.
\newblock


\bibitem[{Vicon Motion Systems Ltd}(2021)]%
        {vicon}
\bibfield{author}{\bibinfo{person}{{Vicon Motion Systems Ltd}}.}
  \bibinfo{year}{2021}\natexlab{}.
\newblock \bibinfo{title}{{Vicon}}.
\newblock \bibinfo{howpublished}{\url{https://www.vicon.com/}}.
\newblock
\newblock
\shownote{Accessed: 2021-04-29}.


\bibitem[Vigdor(2019)]%
        {vigdor2019somebody}
\bibfield{author}{\bibinfo{person}{Neil Vigdor}.}
  \bibinfo{year}{2019}\natexlab{}.
\newblock \showarticletitle{Somebody’s watching: Hackers breach ring home
  security cameras}.
\newblock \bibinfo{journal}{\emph{The New York Times}} (\bibinfo{year}{2019}).
\newblock


\bibitem[Wang et~al\mbox{.}(2014)]%
        {wang2014rf}
\bibfield{author}{\bibinfo{person}{Jue Wang}, \bibinfo{person}{Deepak Vasisht},
  {and} \bibinfo{person}{Dina Katabi}.} \bibinfo{year}{2014}\natexlab{}.
\newblock \showarticletitle{RF-IDraw: Virtual touch screen in the air using RF
  signals}.
\newblock \bibinfo{journal}{\emph{ACM SIGCOMM Computer Communication Review}}
  \bibinfo{volume}{44}, \bibinfo{number}{4} (\bibinfo{year}{2014}),
  \bibinfo{pages}{235--246}.
\newblock


\bibitem[Wang et~al\mbox{.}(2019a)]%
        {wang2019efficient}
\bibfield{author}{\bibinfo{person}{Tianyu Wang}, \bibinfo{person}{Hanying
  Zhao}, {and} \bibinfo{person}{Yuan Shen}.} \bibinfo{year}{2019}\natexlab{a}.
\newblock \showarticletitle{An efficient single-anchor localization method
  using ultra-wide bandwidth systems}.
\newblock \bibinfo{journal}{\emph{Applied Sciences}} \bibinfo{volume}{10},
  \bibinfo{number}{1} (\bibinfo{year}{2019}), \bibinfo{pages}{57:1--17}.
\newblock


\bibitem[Wang et~al\mbox{.}(2019b)]%
        {wang2019high}
\bibfield{author}{\bibinfo{person}{Tianyu Wang}, \bibinfo{person}{Hanying
  Zhao}, {and} \bibinfo{person}{Yuan Shen}.} \bibinfo{year}{2019}\natexlab{b}.
\newblock \showarticletitle{High-accuracy localization using single-anchor
  ultra-wide bandwidth systems}. In \bibinfo{booktitle}{\emph{IEEE/CIC
  International Conference on Communications in China (ICCC)}}.
  \bibinfo{pages}{59--63}.
\newblock


\bibitem[Wang et~al\mbox{.}(2020)]%
        {sensys2020symphony}
\bibfield{author}{\bibinfo{person}{Weiguo Wang}, \bibinfo{person}{Jinming Li},
  \bibinfo{person}{Yuan He}, {and} \bibinfo{person}{Yunhao Liu}.}
  \bibinfo{year}{2020}\natexlab{}.
\newblock \showarticletitle{Symphony: Localizing multiple acoustic sources with
  a single microphone array}. In \bibinfo{booktitle}{\emph{ACM Conference on
  Embedded Networked Sensor Systems (SenSys)}}. \bibinfo{pages}{82--94}.
\newblock


\bibitem[Wang and Dunston(2007)]%
        {wang2007design}
\bibfield{author}{\bibinfo{person}{Xiangyu Wang} {and} \bibinfo{person}{Phillip
  Dunston}.} \bibinfo{year}{2007}\natexlab{}.
\newblock \showarticletitle{Design, strategies, and issues towards an augmented
  reality-based construction training platform}.
\newblock \bibinfo{journal}{\emph{Journal of Information Technology in
  Construction}} \bibinfo{volume}{12}, \bibinfo{number}{25}
  (\bibinfo{year}{2007}), \bibinfo{pages}{363--380}.
\newblock


\bibitem[Wang et~al\mbox{.}(2023)]%
        {imwut2023spectral}
\bibfield{author}{\bibinfo{person}{Yanxiang Wang}, \bibinfo{person}{Jiawei Hu},
  \bibinfo{person}{Hong Jia}, \bibinfo{person}{Wen Hu}, \bibinfo{person}{Mahbub
  Hassan}, \bibinfo{person}{Ashraf Uddin}, \bibinfo{person}{Brano Kusy}, {and}
  \bibinfo{person}{Moustafa Youssef}.} \bibinfo{year}{2023}\natexlab{}.
\newblock \showarticletitle{Spectral-Loc: Indoor localization using light
  spectral information}.
\newblock \bibinfo{journal}{\emph{Proceedings of the ACM on Interactive,
  Mobile, Wearable and Ubiquitous Technologies}} \bibinfo{volume}{7},
  \bibinfo{number}{1} (\bibinfo{year}{2023}), \bibinfo{pages}{1--26}.
\newblock


\bibitem[Xi et~al\mbox{.}(2022)]%
        {xi2022challenges}
\bibfield{author}{\bibinfo{person}{Nannan Xi}, \bibinfo{person}{Juan Chen},
  \bibinfo{person}{Filipe Gama}, \bibinfo{person}{Marc Riar}, {and}
  \bibinfo{person}{Juho Hamari}.} \bibinfo{year}{2022}\natexlab{}.
\newblock \showarticletitle{The challenges of entering the metaverse: An
  experiment on the effect of extended reality on workload}.
\newblock \bibinfo{journal}{\emph{Information Systems Frontiers}}
  (\bibinfo{year}{2022}), \bibinfo{pages}{1--22}.
\newblock


\bibitem[Xie et~al\mbox{.}(2020)]%
        {sensys2020litag}
\bibfield{author}{\bibinfo{person}{Pengjin Xie}, \bibinfo{person}{Lingkun Li},
  \bibinfo{person}{Jiliang Wang}, {and} \bibinfo{person}{Yunhao Liu}.}
  \bibinfo{year}{2020}\natexlab{}.
\newblock \showarticletitle{LiTag: localization and posture estimation with
  passive visible light tags}. In \bibinfo{booktitle}{\emph{ACM Conference on
  Embedded Networked Sensor Systems (SenSys)}}. \bibinfo{pages}{123--135}.
\newblock


\bibitem[Xiong and Jamieson(2013)]%
        {arraytrack}
\bibfield{author}{\bibinfo{person}{Jie Xiong} {and} \bibinfo{person}{Kyle
  Jamieson}.} \bibinfo{year}{2013}\natexlab{}.
\newblock \showarticletitle{{ArrayTrack: A Fine-grained indoor location
  system}}. In \bibinfo{booktitle}{\emph{{USENIX} Symposium on Networked
  Systems Design and Implementation ({NSDI})}}. \bibinfo{pages}{71--84}.
\newblock


\bibitem[Xue et~al\mbox{.}(2021)]%
        {xue2021mmmesh}
\bibfield{author}{\bibinfo{person}{Hongfei Xue}, \bibinfo{person}{Yan Ju},
  \bibinfo{person}{Chenglin Miao}, \bibinfo{person}{Yijiang Wang},
  \bibinfo{person}{Shiyang Wang}, \bibinfo{person}{Aidong Zhang}, {and}
  \bibinfo{person}{Lu Su}.} \bibinfo{year}{2021}\natexlab{}.
\newblock \showarticletitle{mmMesh: Towards 3D real-time dynamic human mesh
  construction using millimeter-wave}. In \bibinfo{booktitle}{\emph{ACM
  International Conference on Mobile Systems, Applications, and Services
  (MobiSys)}}. \bibinfo{pages}{269--282}.
\newblock


\bibitem[Yan et~al\mbox{.}(2021)]%
        {sensys2021curvelight}
\bibfield{author}{\bibinfo{person}{Shangyao Yan}, \bibinfo{person}{Zhimeng
  Yin}, {and} \bibinfo{person}{Guang Tan}.} \bibinfo{year}{2021}\natexlab{}.
\newblock \showarticletitle{CurveLight: An accurate and practical indoor
  positioning system}. In \bibinfo{booktitle}{\emph{ACM Conference on Embedded
  Networked Sensor Systems (SenSys)}}. \bibinfo{pages}{152--164}.
\newblock


\bibitem[Yang et~al\mbox{.}(2022)]%
        {yang2022vuloc}
\bibfield{author}{\bibinfo{person}{Jing Yang}, \bibinfo{person}{BaiShun Dong},
  {and} \bibinfo{person}{Jiliang Wang}.} \bibinfo{year}{2022}\natexlab{}.
\newblock \showarticletitle{VULoc: Accurate UWB localization for countless
  targets without synchronization}.
\newblock \bibinfo{journal}{\emph{Proceedings of the ACM on Interactive,
  Mobile, Wearable and Ubiquitous Technologies}} \bibinfo{volume}{6},
  \bibinfo{number}{3} (\bibinfo{year}{2022}), \bibinfo{pages}{1--25}.
\newblock


\bibitem[Yang et~al\mbox{.}(2014)]%
        {tagoram}
\bibfield{author}{\bibinfo{person}{Lei Yang}, \bibinfo{person}{Yekui Chen},
  \bibinfo{person}{Xiang~Yang Li}, \bibinfo{person}{Chaowei Xiao},
  \bibinfo{person}{Mo Li}, {and} \bibinfo{person}{Yunhao Liu}.}
  \bibinfo{year}{2014}\natexlab{}.
\newblock \showarticletitle{Tagoram: Real-time tracking of mobile {RFID} tags
  to high precision using {COTS} devices}. In \bibinfo{booktitle}{\emph{ACM
  Annual International Conference on Mobile Computing and Networking
  (MobiCom)}}. \bibinfo{pages}{237--248}.
\newblock


\bibitem[Zhang et~al\mbox{.}(2022)]%
        {zhang2022toward}
\bibfield{author}{\bibinfo{person}{Xianan Zhang}, \bibinfo{person}{Lieke Chen},
  \bibinfo{person}{Mingjie Feng}, {and} \bibinfo{person}{Tao Jiang}.}
  \bibinfo{year}{2022}\natexlab{}.
\newblock \showarticletitle{Toward reliable non-line-of-sight localization
  using multipath reflections}.
\newblock \bibinfo{journal}{\emph{Proceedings of the ACM on Interactive,
  Mobile, Wearable and Ubiquitous Technologies}} \bibinfo{volume}{6},
  \bibinfo{number}{1} (\bibinfo{year}{2022}), \bibinfo{pages}{1--25}.
\newblock


\bibitem[Zhao et~al\mbox{.}(2021)]%
        {zhao2021uloc}
\bibfield{author}{\bibinfo{person}{Minghui Zhao}, \bibinfo{person}{Tyler
  Chang}, \bibinfo{person}{Aditya Arun}, \bibinfo{person}{Roshan
  Ayyalasomayajula}, \bibinfo{person}{Chi Zhang}, {and} \bibinfo{person}{Dinesh
  Bharadia}.} \bibinfo{year}{2021}\natexlab{}.
\newblock \showarticletitle{ULoc: Low-power, scalable and cm-accurate UWB-tag
  localization and tracking for indoor applications}.
\newblock \bibinfo{journal}{\emph{Proceedings of the ACM on Interactive,
  Mobile, Wearable and Ubiquitous Technologies}} \bibinfo{volume}{5},
  \bibinfo{number}{3} (\bibinfo{year}{2021}), \bibinfo{pages}{1--31}.
\newblock


\bibitem[Zhong et~al\mbox{.}(2021)]%
        {zhong2021towards}
\bibfield{author}{\bibinfo{person}{Fangwei Zhong}, \bibinfo{person}{Peng Sun},
  \bibinfo{person}{Wenhan Luo}, \bibinfo{person}{Tingyun Yan}, {and}
  \bibinfo{person}{Yizhou Wang}.} \bibinfo{year}{2021}\natexlab{}.
\newblock \showarticletitle{Towards distraction-robust active visual tracking}.
  In \bibinfo{booktitle}{\emph{International Conference on Machine Learning
  (ICML)}}. \bibinfo{pages}{12782--12792}.
\newblock


\bibitem[Zwirello et~al\mbox{.}(2012)]%
        {zwirello2012uwb}
\bibfield{author}{\bibinfo{person}{Lukasz Zwirello}, \bibinfo{person}{Tom
  Schipper}, \bibinfo{person}{Marlene Harter}, {and} \bibinfo{person}{Thomas
  Zwick}.} \bibinfo{year}{2012}\natexlab{}.
\newblock \showarticletitle{UWB localization system for indoor applications:
  Concept, realization and analysis}.
\newblock \bibinfo{journal}{\emph{Journal of Electrical and Computer
  Engineering}}  \bibinfo{volume}{2012} (\bibinfo{year}{2012}),
  \bibinfo{pages}{1--12}.
\newblock


\end{thebibliography}
\end{document}